\documentclass[aps,prb,10pt,onecolumn,superscriptaddress,nofootinbib,showkeys]{revtex4-2}
\usepackage[T1]{fontenc}
\usepackage[utf8]{inputenc}
\usepackage{graphicx}
\usepackage{amsmath,amssymb}
\usepackage{microtype}
\usepackage[hidelinks]{hyperref}

\begin{document}

\title{Topological Optical Frequency Combs}

\author{Rujiang Li}
\email{rujiangli@xidian.edu.cn}
\affiliation{National Key Laboratory of Radar Detection and Sensing, School of Electronic Engineering, Xidian University, Xi'an 710071, China}

\author{Wenfu Zhang}
\email{wfuzhang@opt.ac.cn}
\affiliation{State Key Laboratory of Ultrafast Optical Science and Technology, Xi’an Institute of Optics and Precision Mechanics, Chinese Academy of Sciences, Xi'an 710119, China}
\affiliation{University of Chinese Academy of Sciences, Beijing 100049, China}

\begin{abstract}
Optical frequency combs (OFCs) are revolutionary light sources characterized by discrete and equally spaced spectral lines, and they have found widespread applications in metrology, spectroscopy, and communications. In the early stages, OFCs were realized using mode-locked lasers. With advancements in the fabrication of high-quality factor ($Q$) microresonators and the increasing demand for miniaturized and integrable photonic chips, microresonator-based OFCs, commonly referred to as microcombs, have been developed. Although early studies of microcombs primarily focused on single microresonators or a few resonators, a significant breakthrough occurred in 2021 when it was theoretically predicted that light propagating in the topological edge channel of an array of ring resonators could generate nested frequency combs known as topological OFCs. Since then, the field of topological OFCs has progressed rapidly, with experimental observations made in 2024. This Perspective will introduce the history of OFCs, placing particular emphasis on the emergence and development of topological OFCs, as well as exploring the research challenges and opportunities associated with them.
\end{abstract}

\maketitle

\section{Introduction}

Optical  frequency combs (OFCs) are revolutionary light sources characterized by discrete, equally spaced spectral lines in the frequency domain and highly coherent ultrashort pulse trains in the time domain \cite{RMP75-325}. Over the past five decades, OFCs have evolved from their inception into a rapidly advancing technology, with a key objective being to achieve high performance on a miniaturized and integrable platform \cite{AP2-1,eLight4-19,PI3-R09,nphoton16-95}. Notably, the Nobel Prize in Physics was awarded in 2005 to Theodor W. H\"{a}nsch and John L. Hall for their pioneering contributions to the field of OFCs \cite{RMP78-1297,RMP78-1279}, which served as a significant catalyst for further advancements in this technology. While the principles of OFCs have matured significantly \cite{arxiv,PTRSA}, and their applications have expanded to include optical atomic clocks \cite{nature624-267,nphoton19-400}, photonic microwave generation \cite{nature627-540,nature627-534,nature627-546,nelectron7-1170,AP11-1412,nc14-3467}, optical communications \cite{nphoton16-798,nphoton19-451,PR10-2802,nc17-9319,nc15-7892}, precision spectroscopy \cite{PRX15-011061,nphoton18-1195,SCPMA65-294211}, astronomical spectrograph calibration \cite{nc15-7614,nc15-1466,OE34-8569}, optical computing \cite{nc14-66,nc16-292,eLight5-20,eLight5-10}, laser ranging \cite{sa11-eadt4252,sa11-eads9590,nc13-3280}, and many other areas, intriguing challenges and opportunities remain. These uncharted territories hold the potential to redefine the future of OFCs.

\section{From Mode-Locked Lasers to Microcombs}

\begin{figure}[tbp]
\centering
\includegraphics[width=\linewidth]{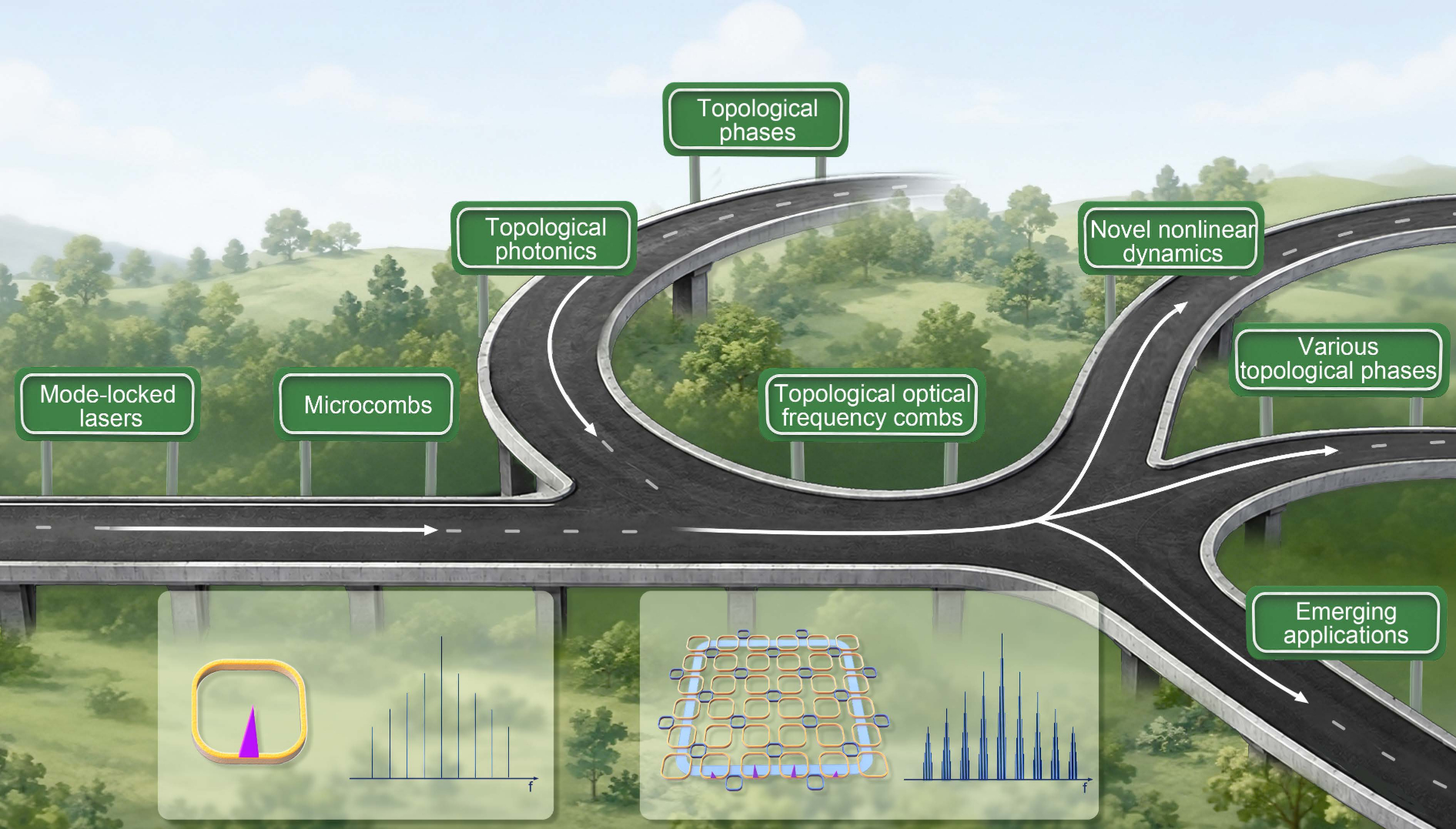}
\caption{Roadmap of topological optical frequency combs (OFCs). In the early stages, OFCs were realized using mode-locked lasers. Then the advent of high-quality factor ($Q$) optical microresonators led to the emergence of microresonator-based OFCs, known as microcombs. Recently, topological photonics, rooted in the introduction of topological phases into photonics, has converged with OFCs, resulting in the emergence of topological OFCs. Future development directions for topological OFCs may include exploring novel nonlinear dynamics, incorporating various types of topological phases, and discovering emerging applications. The inset beneath the ``Microcombs'' guidepost depicts a typical microring resonator structure along with its corresponding frequency comb. Meanwhile, the inset under ``Topological optical frequency combs'' illustrates an array of coupled microring resonators that generate nested frequency combs, where the comb teeth from a single resonator are subdivided into a series of teeth.}
\label{fig_roadmap}
\end{figure}

The original idea of OFCs can be traced back to the 1970s. T. W. H\"{a}nsch independently proposed this notion \cite{hansch}, while Ye.~V. Baklanov and V. P. Chebotayev concurrently introduced similar concepts \cite{AP12-97}. They noted that a regularly spaced train of pulses corresponds to a comb-like spectrum in the frequency domain, which can be utilized for high-resolution laser spectroscopy. Soon after these theoretical proposals, a train of phase-coherent standing-wave light pulses was experimentally realized through multiple reflections of a single laser pulse inside an optical resonator~\cite{PRL38-760}. Later, a train of picosecond light pulses was generated using a synchronously pumped mode-locked dye laser, and the actively controlled mode spectrum provided a means for accurate frequency measurement \cite{PRL40-847}. Since then, mode-locked laser technology has continued to evolve, encompassing a wide range of devices from mode-locked solid-state lasers \cite{OL24-881,PRL82-3568} to mode-locked fiber lasers \cite{OC183-181,OE11-594,OE12-5872,OE10-1404}. However, the generation of OFCs still fundamentally relies on these lasers. As illustrated in Figure \ref{fig_roadmap}, mode-locked lasers were the primary devices responsible for generating OFCs in the early stages. This reliance has led to OFC devices being relatively bulky \cite{RMP75-325, nphoton16-95}.

The development of OFCs has entered a new stage with the advent of optical microresonators \cite{PI3-R09,AP2-1,eLight4-19,science332-555,science361-eaan8083,nphoton16-95,optica10-977,aop18-86,npjnano1-26,aplphoton7-100901,PQE86-100437,arxiv}. Thanks to the rapid advancements in micro/nanofabrication technologies, the fabrication of high-quality factor ($Q$) resonators has become feasible \cite{OL47-1855,nc16-8878,optica11-1397,COL20-032201}. An early landmark work in this context was the demonstration of whispering gallery mode (WGM) microresonators reported in 1989, which achieved a $Q$ value of $Q \geq 10^8$ \cite{PLA137-393}. In high-$Q$ microresonators, the optical field can be significantly enhanced, dramatically lowering the power threshold required for nonlinear optical effects, such as the parametric process of four-wave mixing (FWM) in microresonators made from materials exhibiting intrinsic Kerr nonlinearity \cite{SCPMA65-104201}. Furthermore, due to their unprecedentedly small sizes, optical microresonators provide opportunities for creating a low-power, compact, and integrable platform for generating OFCs \cite{nc15-4192,nature605-457,LPR20-e01659,nc15-7030,nc16-4829}.

In 2004, Kerr-nonlinearity-induced optical parametric oscillation (OPO) was experimentally realized in high-$Q$ WGM resonators, where only several discrete frequencies play a role in the process \cite{PRL93-083904,PRL93-243905}. Later, in 2007, T. J. Kippenberg’s group experimentally demonstrated the generation of OFCs from a silica toroidal microcavity through cascaded FWM parametric oscillation \cite{nature450-1214}. Since then, OFCs have been generated in various types of microresonators, including $\mathrm{CaF}_{2}$ WGM resonators \cite{PRL101-093902,OL34-878}, fiber cavities \cite{PRL102-193902,LSA12-33,nc15-55}, silica microspheres \cite{OE17-16209}, silicon nitride resonators \cite{OL36-3398,OL37-875,nphoton5-770,nphoton18-294,nature646-843}, and $\mathrm{MgF}_{2}$ resonators \cite{OL36-2290,nc4-1345,LPR18-2301329}. These microresonator-based OFCs, particularly those generated via the Kerr effect, are commonly referred to as microcombs \cite{PI3-R09,AP2-1,eLight4-19,PQE86-100437}. The era of microcombs began following the development of mode-locked lasers, as illustrated in Figure \ref{fig_roadmap}.

Ideally, OFCs should exhibit equidistant frequency spacing, and their spectral components should be coherent with minimal phase noise. However, these conditions are not always met in experiments \cite{nphoton5-770,nphoton6-480,nc14-4590,LSA13-66}. According to the microcomb theory, the generation of OFCs is typically described by the Lugiato-Lefever equation (LLE) \cite{PRL58-2209}.  Depending on the pump power, cavity detuning, and dispersion, the LLE can support homogeneous continuous-wave states, periodic Turing rolls arising from modulational instability, localized solitons, and time-dependent breathing and chaotic states \cite{PRA89-063814}. These soliton solutions, known as dissipative Kerr solitons (DKSs), are sustained by the balance between external pumping and loss, as well as the balance between group-velocity dispersion (GVD) and Kerr nonlinearity \cite{OC91-401,PI3-R09,science361-eaan8083,OL18-601}. The discovery of DKSs provided a route to achieving mode-locked microcombs with high coherence \cite{OL36-2845,PRL136-063801,nc14-1802}.

Historically, temporal DKSs were first experimentally observed in a passive ring cavity made of standard optical fibers \cite{nphoton4-471}. Due to the thermally induced resonance shifts encountered during the experimental generation of soliton states, the realization of soliton microcombs, where microresonator-based OFCs operate in the state of temporal DKSs, was not demonstrated until 2014 \cite{nphoton8-145}. In that work, a crystalline $\mathrm{MgF}_{2}$ resonator was used. Following this landmark study, soliton microcombs have been successively demonstrated in a wide variety of microresonators, including silica \cite{optica2-1078,nc14-169,nc15-1661}, silicon nitride \cite{science351-357,nc10-680,nphoton19-630,nature562-401,LSA15-370,nphoton18-632,optica10-650}, aluminum nitride \cite{OL43-4366,OL47-746,PR11-A10}, AlGaAs \cite{optica3-823,nphoton18-625,OL48-3853}, silicon \cite{ncommun6-6299,optica3-854}, chalcogenide glass (GeSbS) \cite{LPR17-2200219,nc16-10133}, gallium phosphide \cite{optica11-1454}, $\mathrm{LiTaO}_{3}$ \cite{nature629-784,PR13-1955}, and $\mathrm{LiNbO}_{3}$ \cite{nc16-2389,LSA14-270,LSA13-225,nc15-3921,eLight5-15,LPR18-2300627}. Meanwhile, numerous phenomena related to microcomb dynamics have been revealed, including flat-top solitons known as platicons in the normal-dispersion regime \cite{nphoton9-594,OE23-7713,nc13-1771,CP6-303,nc13-3134}, dark-bright soliton pairs \cite{PRL128-033901}, soliton crystals \cite{nphoton11-671,nc12-3179,nphys15-1071,OE30-13690,LSA11-341,LSA13-251}, and discrete time crystals \cite{nc13-848,LPR19-2500257}.

The early studies of soliton microcombs primarily focused on single microresonators. In Figure \ref{fig_roadmap}, the inset beneath the ``Microcombs'' guidepost depicts a typical microring resonator structure along with its corresponding frequency comb. In recent years, researchers have shifted their attention to multiple resonators \cite{science383-1080,optica10-279,nc16-4780}. In photonic dimers, which consist of two coupled microresonators, novel dissipative solitons and emergent nonlinear phenomena that are unattainable in single resonator systems have been discovered \cite{nphoton15-305,nphys17-604,sa8-eabm6982,CP9-206,PRL134-123801,optica11-940}. From an application perspective, utilizing mutually coupled cavities instead of a traditional single cavity can enhance the power efficiency of soliton generation \cite{nphoton13-616,nphoton17-992,LSA15-185}. Meanwhile, multicolour pulse pairs can be generated in multiple coupled resonators \cite{nphoton17-977,LSA15-166}. Considering the recent study of symmetry breaking in nonlinear microresonator arrays \cite{PR12-2376,OL51-1649}, it is predictable that the introduction of a spatial dimension will often bring about new fundamental physics as well as application insights. This assertion also underscores P. W. Anderson's statement that ``more is different'' \cite{anderson}. Thus, the shift from single microresonators to resonator arrays opens a new chapter for OFCs.

\section{Topological Photonics}

When the study of soliton microcombs extends into resonator arrays, two seemingly unrelated fields, OFCs and topological photonics, naturally converge. Topological photonics is fundamentally rooted in the introduction of topological phases into photonics. As shown in Figure \ref{fig_roadmap}, where the guidepost for ``Topological phases'' branches off from a separate path, the concept of topological phases originates from condensed matter physics. It defines distinct states of matter characterized by global properties that remain invariant under continuous transformations, with their behavior governed by topological invariants. Following the seminal discovery of the quantum Hall effect and the TKNN invariant \cite{PRL49-405}, substantial research efforts have been dedicated to exploring various types of topological phases \cite{RMP82-3045,RMP83-1057,NRM7-196}. Among these, topological insulators stand out as a prominent example, functioning as conventional insulators in the bulk while exhibiting conductive properties on their surfaces \cite{PRL95-226801,science318-766}. This conductivity arises from topologically protected edge states that are robust against disorder and defects. As research into topological phases has flourished, a noteworthy milestone was the awarding of the 2016 Nobel Prize in Physics to David J. Thouless, F. Duncan M. Haldane, and J. Michael Kosterlitz for their theoretical discoveries regarding topological phase transitions and topological phases of matter \cite{RMP89-040502,RMP89-040501}.

In 2008, F. D. M. Haldane and S. Raghu pioneeringly introduced topological phases into the realm of photonics~\cite{PRL100-013904}. They proposed that the use of magneto-optic materials could break time-reversal symmetry, thus enabling the emulation of the quantum Hall effect in photonic crystals and achieving unidirectional photonic modes resembling chiral edge states. Subsequently, Zheng Wang et al. from MIT theoretically proposed a more general scheme that eliminated the need to construct Dirac points \cite{PRL100-013905}. In 2009, this theoretical proposal was experimentally verified in a magneto-optical photonic crystal fabricated in the microwave regime, where unidirectional backscattering-immune topological electromagnetic states were observed \cite{nature461-772}. These pioneering works paved the way for the exploration of novel topological phases in photonics, and the immunity of topological edge states to local deformations and disorders is crucial for various promising applications \cite{RMP91-015006, LSA9-1, APR,nc15-931}. Since then, the field of topological photonics has developed rapidly, as illustrated in Figure \ref{fig_roadmap}.

Nowadays, various topological photonic structures have been proposed \cite{NRP8-327}. Several typical examples include resonator arrays \cite{nphys7-907,nphoton7-1001,np6-782,nmater23-928}, metamaterials \cite{nmater12-233,nmater15-542,nc5-5782,CR123-7585}, optical waveguide arrays \cite{nature496-196,nc17-3020,nc17-966,nc15-9311}, and photonic crystals \cite{nmater16-298,nphys14-140,LPR16-2100300,PNAS121-e2411793121,nc13-6738}. Among these types of topological photonic structures, resonator arrays have direct applications in silicon photonics and are closely related to the topological OFCs that will be discussed later. In 2011, Mohammad Hafezi et al. theoretically proposed that a quantum spin Hall-like photonic topological insulator could be constructed using a network of coupled resonator optical waveguides, specifically an array composed of optical ring microresonators \cite{nphys7-907}. The degenerate clockwise and counterclockwise modes within a single resonator can be treated as two components of a pseudo-spin. Although the time-reversal symmetry of the system is not explicitly broken, robust edge state transport can be anticipated because each spin component experiences a spin-dependent magnetic field and exhibits quantum Hall-like behaviors. This theoretical proposal was experimentally realized by the same group in 2013 \cite{nphoton7-1001}. Interestingly, while both studies were conducted in the linear regime, i.e., neglecting the ambient medium-mediated interactions between photons, the authors suggest that intriguing avenues could emerge with the inclusion of interactions between photons. Such predictions foreshadow the subsequent convergence of OFCs and topological photonics.

With the gradual maturation of topological photonics in the linear regime, there has been a growing interest in exploring the new features that emerge when nonlinear optical effects are introduced into topological lattices~\cite{APR7-021306,nphys20-905,PRA113-053506,CP5-275,PRB105-L201111,AP12-2291}. Traditionally, a wide range of nonlinear optical phenomena has been studied using optical waveguide arrays \cite{RMP83-247}. On one hand, by treating the propagation direction of optical waveguide arrays as the temporal dimension, a direct equivalence can be established between the paraxial wave equation and the Schr\"{o}dinger equation. On the other hand, due to the rapid development of laser-writing techniques, various types of optical waveguide arrays, including helical waveguides, can now be fabricated. For these reasons, novel types of nonlinear states, including solitons, have been discovered in nonlinear topological lattices based on optical waveguide arrays \cite{PRA90-023813,PRA94-021801,PRL128-093901,PRL117-143901,PRX11-041057,PRL111-243905,science368-856,nphys17-995,LSA13-264,LSA12-194}. Additionally, it has been proposed that nonlinearity can be utilized to manipulate topological edge states \cite{LSA9-147,science372-72,LSA10-164,nanophotonics14-769,PRL127-184101,nature596-63,PRL128-154101,PRL128-113901,ncommun13-5997,
nphys19-420,CP8-360}.

Compared to the extensive studies in nonlinear topological photonics based on optical waveguide arrays, studies of resonator arrays, another prominent example of the earliest proposed platforms, remain largely confined to the linear regime of topological photonics. Nevertheless, since 2018, topological edge states in resonator arrays have increasingly been utilized for the generation of correlated photon pairs \cite{nature561-502,nphoton16-248} and indistinguishable photon pairs \cite{nphoton15-542} through spontaneous FWM. Meanwhile, it has been reported that the third-harmonic signal can be significantly enhanced by utilizing the edge state of a topologically non-trivial zigzag array of dielectric nanoresonators \cite{nnano14-2}. This study brings concepts of nonlinear topological photonics into the realm of nanoscience.

\section{Topological Optical Frequency Combs}

In 2017, the generation of OFCs in cascaded parity-time (PT)-symmetric Aubry-Andr\'e-Harper lattices was theoretically proposed \cite{OL42-5174}. A striking breakthrough occurred in 2021 when Hafezi's group theoretically proposed the formation of nested OFCs and DKSs in a topological microresonator array \cite{nphys17-1169}. They pointed out that the OFCs generated in the topological microresonator array exhibit a nested spectral structure, in which the comb teeth of the super-ring resonator are embedded within those of a single isolated resonator. In their work, these OFCs were referred to as topological OFCs and the DKSs as nested solitons. In a News \& Views article on this work, Vittorio Peano vividly described these topological OFCs as Matryoshka frequency combs \cite{nphys17-1078}. This work marked a major milestone in the development of topological OFCs, as illustrated in Figure \ref{fig_roadmap}, where the guidepost labeled ``Topological optical frequency combs'' appears after the crossover of ``Microcombs'' and ``Topological photonics''.

More specifically, S. Mittal et al. constructed a two-dimensional (2D) array of coupled microring resonators~\cite{nphys17-1169}. Related structures had been previously studied in the linear regime by the same group \cite{nphys7-907,nphoton7-1001}. As discussed earlier, although the system is time-reversal symmetric and exhibits two components of pseudo-spin corresponding to the clockwise and anticlockwise modes in a single resonator, the coupling between these two components can be neglected, and each pseudo-spin component can be excited independently. In this context, for one of the pseudo-spin components, the system simulates the anomalous quantum Hall effect, with topological edge states emerging as one pseudo-spin component circulates around the lattice boundary. Consequently, the perimeter of the resonator array can be perceived as a travelling-wave super-ring resonator. To illustrate this, Figure \ref{fig_roadmap} features an array of coupled microring resonators in the inset under the guidepost ``Topological optical frequency combs'', where a blue-colored super-ring resonator is formed due to the circulation of topological edge states. The nonlinear dynamics of these topological edge states in the super-ring resonator closely resemble those of excitations in a single nonlinear resonator. This self-similarity forms the basis of the topological OFCs and nested solitons, distinguishing them sharply from all previously studied OFCs.

Mittal et al. showed that when the microresonator array is externally pumped by a continuous wave, varying the pump power and detuning frequency leads the super-ring resonator to exhibit three distinct regimes of nonlinear dynamics: chaotic regions, Turing rolls, and solitons \cite{nphys17-1169}. The presence of these regimes aligns well with the behavior observed in single-resonator-based OFCs \cite{PI3-R09,arxiv}. In the regime of nested solitons, the temporal pulses produced by the microresonator array are separated by the single-ring round-trip time and are modulated by a series of super-soliton pulses, which are separated by the round-trip time of the edge mode super-ring. Consequently, the output spectrum displays two sets of nested comb teeth. The comb teeth generated by the super-ring resonator correspond to each tooth of the frequency comb produced by a single resonator. In other words, the comb teeth from a single resonator are subdivided into a series of teeth in the frequency comb generated by the super-ring resonator. The nested frequency combs are also depicted in the inset beneath the guidepost ``Topological optical frequency combs'' in Figure \ref{fig_roadmap}.

Since the nested solitons are inherited from the linear topological edge states, Mittal et al. claimed that they are topologically protected in the sense that a nested soliton could bypass a strongly detuned edge resonator while preserving phase coherence, without observable reflection or scattering into the bulk \cite{nphys17-1169}. Topological edge transport provides a mechanism for defect tolerance that is absent in conventional single-resonator-based OFCs. However, the existence and dynamical stability of nested solitons also depend on the pumping conditions, dispersion, and losses, and are not guaranteed by the underlying linear topology alone. Additionally, Mittal et al. found that a single nested soliton achieves a mode efficiency of over $50\%$, an order of magnitude higher than that of single-ring frequency combs in the single-soliton regime \cite{nphys17-1169}.
Thus, topological OFCs offer a route to improved tolerance to fabrication defects and higher mode efficiency.

This Perspective mainly focuses on comb generation in photonic structures with nontrivial band topology. Very recently, Nicolas Englebert et al. demonstrated the realization of topological OFCs based on the quadratic nonlinearity provided by lithium niobate \cite{arxiv-OFC}. Due to the balance between dispersion and quadratic nonlinearity, a dark pulse consisting of topological phase defects that separate two $\pi$-out-of-phase continuous-wave solutions is supported \cite{nphoton20-616}. Considering the formation of domain walls in the fast time dimension, they referred to this type of solitons as topological solitons and designated the OFCs as topological soliton frequency combs or topological frequency combs. However, Nicolas Englebert et al. still conventionally used a single resonator, and there are no topological edge states circulating around the perimeter of a resonator array. A more in-depth discussion of the topological OFCs proposed by Nicolas Englebert et al. is beyond the scope of this Perspective.

The generation of nested topological OFCs was experimentally demonstrated by Flower et al. in 2024 \cite{science384-1356}.
The difficulties in observing topological OFCs may arise from two main aspects. First, while the fabrication of single nonlinear resonators has matured, creating large arrays of microresonators remains a challenge. The previous demonstration of such resonator arrays was focused on the linear regime, where the nonlinear properties and dispersions of the resonators were not a primary concern \cite{nphoton7-1001}. The nonlinear regime requires more careful design and optimization of resonator parameters. Despite these challenges, Christopher J. Flower et al. successfully fabricated a 2D lattice containing over 100 ring resonators made from silicon nitride (SiN) \cite{science384-1356}. The second challenge lies in the experimental implementation of external pumping. Conventionally, to generate OFCs, researchers typically use continuous-wave lasers, as they can output high-power, single-frequency waves. However, the detrimental thermal effects associated with continuous-wave lasers can potentially damage the photonic chip. To address this issue, Christopher J. Flower et al. employed a pulsed tunable laser that mitigates these thermal effects \cite{science384-1356}. Additionally, their pulsed laser has a relatively long pulse duration and low repetition rate, ensuring that its output functions as a high-power quasi-continuous wave, meeting the requirements for OFC generation.

By overcoming the aforementioned difficulties, Flower et al. successfully observed the topological OFCs in their experiments \cite{science384-1356}. Specifically, to demonstrate the nested structure of the topological OFCs, they measured the comb output using an ultra-high-resolution optical spectrum analyzer and discovered another set of well-resolved modes within each individual comb tooth of the low-resolution comb spectrum \cite{science384-1356}. Furthermore, to confirm that the topological OFCs are indeed inherited from the linear topological edge states, they conducted direct imaging of the generated comb \cite{science384-1356}. They observed that the comb light is confined to the edge of the lattice and travels unidirectionally from the input to the output port without noticeable scattering.

The theoretical proposal by Mittal et al. \cite{nphys17-1169} and the experimental demonstration by Flower et al. \cite{science384-1356} have rapidly ignited an explosive growth in the field of topological OFCs. As this area continues to evolve swiftly, the following paragraphs of this Perspective will focus on several key works that stem from the foundational studies and explore future research directions.

In 2023, Aleksandr Tusnin et al. derived an effective $\left( N+1 \right)$D coupled-mode equation for the general case of nonlinear dynamics in an arbitrary $N$-dimensional lattice of resonators \cite{cp6-317}. In the same year, Jiang et al. theoretically proposed the generation of DKS combs in a valley-Hall photonic crystal resonator \cite{PRB108-205421}. They noted that complex coupled-ring arrays can pose challenges for chip miniaturization and optical modulation, while also complicating the modeling of nonlinear dynamics. Their approach uses a compact single-resonator architecture in which circulating valley kink states form the resonator modes that support comb generation. The nonlinear dynamics can be described by a single-resonator LLE, and the generated combs do not exhibit the nested spectral structure characteristic of the coupled-ring-array scheme.
In 2024, the same group theoretically proposed the generation of quantum OFCs with high-dimensional frequency entanglement through spontaneous FWM in valley-Hall topological resonators \cite{AQT}. In 2025, they experimentally demonstrated robust on-chip topological transport of quantum OFCs and DKS combs through valley photonic crystal waveguides, with both types of combs generated in a separate conventional microresonator \cite{PR13-163}. The first two studies by Jiang et al. therefore concern comb generation in topological resonators, whereas their 2025 study concerns the topological transport of externally generated combs. In particular, the latter two studies suggest that the convergence of OFCs and topological photonics has begun to be realized in the quantum regime.

In 2025, Lei Huang et al. proposed the generation of OFCs and nested temporal solitons in hyperbolic photonic topological insulators \cite{AP12-219}. In two-dimensional Euclidean edge-state arrays, a substantial number of bulk resonators are required to support the comb-generating edge channel, limiting resonator utilization \cite{AP12-219}.
Hyperbolic lattices are regular tessellations in non-Euclidean space characterized by constant negative curvature.
Compared with Euclidean lattices, hyperbolic lattices have a larger proportion of boundary sites, allowing a greater fraction of resonators to participate in edge-state comb generation. Using this property, Lei Huang et al. showed that hyperbolic topological OFCs exhibit smaller frequency spacing than their Euclidean counterparts with the same number of site resonators \cite{AP12-219}.
Consequently, fewer site resonators are required to achieve the same frequency spacing, improving resonator utilization.
However, adding lattice layers leads to exponential growth in the number of sites, which in turn increases spatial crowding near the boundary, thereby posing challenges for large-scale implementation. Moreover, the reported hyperbolic OFCs show no significant bandwidth advantage over their Euclidean counterparts \cite{AP12-219}.
This work paves the way for exploring topological OFCs in various types of topological lattices.

S. D. Hashemi and S. Mittal conducted several studies following the seminal theoretical work on topological OFCs. In 2024, they designed a 2D array of strongly coupled ring resonators exhibiting an anomalous Floquet topological phase and theoretically demonstrated Floquet topological solitons \cite{nc15-9642}. Unlike the weakly coupled arrays producing nested combs \cite{nphys17-1169,science384-1356}, this Floquet scheme operates in the strong-coupling regime. By pumping the edge states of the resonator array, they obtained regularly spaced commensurate combs and, for a different choice of coupling parameters, incommensurate combs whose lines are not equidistant but remain phase-locked. This spectral flexibility broadens the range of accessible comb states. However, the nonuniform spacing of the incommensurate states limits their direct use in applications requiring equidistant comb lines \cite{nc15-9642}.
In 2025, they theoretically proposed nested soliton combs in a one-dimensional periodic resonator array with non-Hermitian dissipative couplings \cite{sa11-eadu6554}. Here, the relevant topology is associated with the winding of the complex band spectrum. By engineering the dissipative couplings and hopping phases, they controlled the dispersion and dissipation of the supermodes, enabling both the number and spacing of comb lines to be tuned. This approach offers post-fabrication spectral reconfigurability, although the engineered dissipation introduces additional optical loss and some coupling configurations can require substantial pump power \cite{sa11-eadu6554}.

\section{Challenges and Future Directions}

Looking ahead, there remain numerous challenges and opportunities in the development of topological OFCs. As discussed earlier, many phenomena related to microcomb dynamics have been unveiled, but current studies of topological OFCs are still primarily focused on the formation of bright pulses, which represent a fundamental class of localized solutions of the (coupled) LLE. Thus, the most urgent need is to explore novel nonlinear dynamics within topological OFCs. As illustrated in the roadmap of topological OFCs in Figure \ref{fig_roadmap}, this area of development is represented as a branch off the main path labeled ``Topological optical frequency combs''. In our view, the introduction of topological OFCs not only integrates the concept of topological phases into OFCs, merging the two fields of topological photonics and OFCs, but also introduces spatial dimensions into conventional OFC systems. The transition from single-resonator-based OFCs to resonator-array-based OFCs is likely to yield new types of nonlinear dynamics, exemplified by the work conducted by S. D. Hashemi and S. Mittal in 2025 \cite{sa11-eadu6554}. From our perspective, studying platicons in the normal dispersion regime is particularly intriguing, as platicons represent another important class of solutions to the LLE \cite{nphoton9-594,OE23-7713,nc13-1771}. In a recent preprint, Hashemi et al. theoretically demonstrated the formation of collective topological dark solitons, platicons, and bright-like solitons in two-dimensional Kerr resonator arrays with normal dispersion \cite{arxiv2}.
Additionally, since nested topological OFCs feature two sets of frequency combs, one associated with the super-ring resonator and the other with the single resonator, it remains an open question whether these two sets of frequency combs can be manipulated independently. Notably, the original work by S. Mittal et al. considered the group velocity dispersion (GVD) of single resonators, while the topological edge states circulating within the super-ring resonator exhibit approximately linear dispersion \cite{nphys17-1169}. Therefore, it would be fascinating to investigate the formation of novel types of DKSs under the combined influence of dispersions from both single and super-ring resonators. Lastly, as novel types of solitons emerge through the balance between nonlinearity and higher-order dispersions \cite{nphoton17-937,optica10-1452,nphoton17-943,nc13-4764,CP6-144,OL50-2073}, exploring these soliton types within the platform of topological OFCs would also be of great interest \cite{OL50-4262,OL47-2438}.

The second research direction involves introducing various types of topological phases into the system of OFCs and studying topological OFCs within different topological lattice models. This direction is also depicted in Figure~\ref{fig_roadmap} as a branch off the main path labeled ``Topological optical frequency combs''. Numerous types of topological phases have been discovered, many of which have been realized in photonics \cite{nature618-687,science383-eadf9621}. Furthermore, topological phases have been extended to non-Hermitian systems \cite{FOP20-44204,NMI6-904,PRB113-085413,PRL130-157201} and non-Euclidean spaces \cite{nc15-1647,PRL129-246402,nc13-2937,nc15-2293}. In summary, the field of linear topological photonics has reached a high level of maturity. Therefore, it is both intriguing and necessary to investigate topological phases in nonlinear systems using the platform of topological OFCs. The work conducted by Lei Huang et al. is a typical example of exploring topological OFCs in non-Euclidean curved spaces \cite{AP12-219}.
In the non-Hermitian scheme of Hashemi and Mittal, topological windings of the complex band spectrum provide a means to engineer both the dispersion and dissipation of the resonator array \cite{sa11-eadu6554}.
This opens up competitive avenues for further exploration of other types of topological phases. In particular, the question of whether higher-order topological phases \cite{nphys17-995,LSA10-164,science357-61,PRL120-026801,CP8-451,CSF207-118044} can be integrated with OFCs to enhance their performance is a compelling one. Additionally, since the Haldane model, which requires breaking time-reversal symmetry, has been experimentally realized in resonator arrays \cite{nphys17-704}, it would be fascinating to develop a topological OFC system that explicitly breaks time-reversal symmetry. Furthermore, it would be particularly interesting to explore whether external pumping can spontaneously induce a topological phase transition \cite{science370-701}, thereby supporting the output of frequency combs in the form of topological solitons. Lastly, the definition of topological invariants in systems of topological OFCs is certainly an important research area \cite{nc13-3379,nc16-422,nphys20-1164,PRB113-214310,nc17-6718}.

Besides the two aforementioned research directions related to fundamental physics, the third direction focuses on exploring the emerging applications of topological OFCs, as illustrated by the third branch off the main path labeled ``Topological optical frequency combs''. The features of topological OFCs extend beyond the conventional definition of OFCs, which typically requires the discrete frequency comb teeth to be equally spaced. Nonetheless, topological OFCs may find their own intriguing application scenarios. For instance, as noted by S. Mittal et al. \cite{nphys17-1169}, a single nested soliton can achieve significantly higher mode efficiency compared to conventional single-resonator-based single-soliton combs, making topological OFCs particularly beneficial for miniaturized and integrable OFC chips, where low power consumption is often required. Additionally, due to their comb-in-a-comb structure, nested topological OFCs can offer small frequency intervals that may be favorable for high-density wavelength division multiplexing and high-precision metrology, where a greater number of optical carriers and superior spectral resolution are essential. Most importantly, topological edge transport could improve the tolerance of OFC chips to fabrication defects, although the achievable tolerance should be assessed together with the requirements for soliton formation and stability \cite{NRP8-55}.

\section{Conclusion}

In conclusion, as illustrated by the roadmap in Figure \ref{fig_roadmap}, starting from conventional mode-locked lasers and single-resonator-based microcombs, the development of OFCs has entered a new era with the introduction of topological phases. We can anticipate that topological OFCs will experience explosive growth in the near future.
Finally, real-space topological textures, such as skyrmions, may offer another direction for OFC research. Karnieli et al. theoretically proposed using spatially engineered three-wave mixing in nonlinear photonic crystals to emulate effective skyrmion textures \cite{NC12-1092}. These textures induce beam deflection during frequency conversion, with the deflection direction controlled by the orbital angular momentum of the pump. The authors suggested that this mechanism could be used to control quantum frequency combs. This provides a starting point for exploring the use of real-space topological textures to manipulate OFCs through nonlinear frequency conversion and controllable beam deflection, while their potential for enabling coherent OFC generation remains an open question.

\section*{Funding}
R.L. was supported by the National Key Research and Development Program of China (Grant No. 2022YFA1404902), National Natural Science Foundation
of China (Grant No. 12104353), Natural Science Basic Research Program of Shaanxi (Program No. 2026JC-YBMS-0036), and Fundamental Research Funds for the Central Universities (Grant No. QTZX25086). W.Z. was supported by the National Natural Science Foundation of China (Grant No. 62525509).

\section*{Acknowledgments}
R.L. is deeply grateful to Ying Liu for her support.


\begin{thebibliography}{999}
\bibitem{RMP75-325}
Cundiff, S.T.; Ye, J. Colloquium: Femtosecond optical frequency combs. \emph{Rev. Mod. Phys.} \textbf{2003}, \emph{75}, 325--342.

\bibitem{AP2-1}
Wang, W.; Wang, L.; Zhang, W. Advances in soliton microcomb generation. \emph{Adv. Photonics} \textbf{2020}, \emph{2}, 034001.

\bibitem{eLight4-19}
Yao, B.C.; Wang, W.T.; Xie, Z.D.; et al. Interdisciplinary advances in microcombs: Bridging physics and information technology. \emph{eLight} \textbf{2024}, \emph{4}, 19.

\bibitem{PI3-R09}
Shu, H.; Shen, B.; Chang, H.; et al. Microcomb technology: From principles to applications. \emph{Photonics Insights} \textbf{2024}, \emph{3}, R09.

\bibitem{nphoton16-95}
Chang, L.; Liu, S.; Bowers, J.E. Integrated optical frequency comb technologies. \emph{Nat. Photonics} \textbf{2022}, \emph{16}, 95--108.

\bibitem{RMP78-1297}
Hänsch, T.W. Nobel Lecture: Passion for precision. \emph{Rev. Mod. Phys.} \textbf{2006}, \emph{78}, 1297--1309.

\bibitem{RMP78-1279}
Hall, J.L. Defining and measuring optical frequencies. \emph{Rev. Mod. Phys.} \textbf{2006}, \emph{78}, 1279--1295.

\bibitem{arxiv}
Herr, T.; Tikan, A.; Kippenberg, T.J. Frequency Combs and Coherent Dissipative Structures in Nonlinear Optical Microresonators. \emph{arXiv} \textbf{2026}, arXiv:2604.05897.

\bibitem{PTRSA}
Lugiato, L.A.; Prati, F.; Gorodetsky, M.L.; et al. From the Lugiato--Lefever equation to microresonator-based soliton Kerr frequency combs. \emph{Phil. Trans. R. Soc. A} \textbf{2018}, \emph{376}, 20180113.

\bibitem{nature624-267}
Moille, G.; Stone, J.; Chojnacky, M.; et al. Kerr-induced synchronization of a cavity soliton to an optical reference. \emph{Nature} \textbf{2023}, \emph{624}, 267--274.

\bibitem{nphoton19-400}
Wu, K.; O'Malley, N.P.; Fatema, S.; et al. Vernier microcombs for integrated optical atomic clocks. \emph{Nat. Photonics} \textbf{2025}, \emph{19}, 400--406.

\bibitem{nature627-540}
Sun, S.; Wang, B.; Liu, K.; et al. Integrated optical frequency division for microwave and mmWave generation. \emph{Nature} \textbf{2024}, \emph{627}, 540--545.

\bibitem{nature627-534}
Kudelin, I.; Groman, W.; Ji, Q.X.; et al. Photonic chip-based low-noise microwave oscillator. \emph{Nature} \textbf{2024}, \emph{627}, 534--539.

\bibitem{nature627-546}
Zhao, Y.; Jang, J.K.; Beals, G.J.; et al. All-optical frequency division on-chip using a single laser. \emph{Nature} \textbf{2024}, \emph{627}, 546--551.

\bibitem{nelectron7-1170}
Kudelin, I.; Shirmohammadi, P.; Groman, W.; et al. An optoelectronic microwave synthesizer with frequency tunability and low phase noise. \emph{Nat. Electron.} \textbf{2024}, \emph{7}, 1170--1175.

\bibitem{AP11-1412}
Niu, R.; Hua, T.P.; Shen, Z.; et al. Ultralow-Noise K-Band Soliton Microwave Oscillator Using Optical Frequency Division. \emph{ACS Photonics} \textbf{2024}, \emph{11}, 1412--1418.

\bibitem{nc14-3467}
He, Y.; Lopez-Rios, R.; Javid, U.A.; et al. High-speed tunable microwave-rate soliton microcomb. \emph{Nat. Commun.} \textbf{2023}, \emph{14}, 3467.

\bibitem{nphoton16-798}
Jørgensen, A.A.; Kong, D.; Henriksen, M.R.; et al. Petabit-per-second data transmission using a chip-scale microcomb ring resonator source. \emph{Nat. Photonics} \textbf{2022}, \emph{16}, 798--802.

\bibitem{nphoton19-451}
Corcoran, B.; Mitchell, A.; Morandotti, R.; et al. Optical microcombs for ultrahigh-bandwidth communications. \emph{Nat. Photonics} \textbf{2025}, \emph{19}, 451--462.

\bibitem{PR10-2802}
Shao, W.; Wang, Y.; Jia, S.; et al. Terabit FSO communication based on a soliton microcomb. \emph{Photonics Res.} \textbf{2022}, \emph{10}, 2802--2808.

\bibitem{nc17-9319}
Wang, F.X.; Zheng, S.T.; Huang, L.; et al. Microcomb-driven large-scale fully connected quantum network. \emph{Nat. Commun.} \textbf{2026}, \emph{17}, 9319.

\bibitem{nc15-7892}
Zhang, X.; Zhou, Z.; Guo, Y.; et al. High-coherence parallelization in integrated photonics. \emph{Nat. Commun.} \textbf{2024}, \emph{15}, 7892.

\bibitem{PRX15-011061}
Wang, Z.; Wang, Y.; Shi, B.; et al. Rhythmic Soliton Interactions for Integrated Dual-Microcomb Spectroscopy. \emph{Phys. Rev. X} \textbf{2025}, \emph{15}, 011061.

\bibitem{nphoton18-1195}
Han, J.J.; Zhong, W.; Zhao, R.C.; et al. Dual-comb spectroscopy over a 100 km open-air path. \emph{Nat. Photonics} \textbf{2024}, \emph{18}, 1195--1202.

\bibitem{SCPMA65-294211}
Wang, Y.; Wang, Z.; Wang, X.; et al. Scanning dual-microcomb spectroscopy. \emph{Sci. China Phys. Mech. Astron.} \textbf{2022}, \emph{65}, 294211.

\bibitem{nc15-7614}
Ludwig, M.; Ayhan, F.; Schmidt, T.M.; et al. Ultraviolet astronomical spectrograph calibration with laser frequency combs from nanophotonic lithium niobate waveguides. \emph{Nat. Commun.} \textbf{2024}, \emph{15}, 7614.

\bibitem{nc15-1466}
Cheng, Y.S.; Dadi, K.; Mitchell, T.; et al. Continuous ultraviolet to blue-green astrocomb. \emph{Nat. Commun.} \textbf{2024}, \emph{15}, 1466.

\bibitem{OE34-8569}
Cheng, Y.S.; Dadi, K.; Newman, W.; et al. Comb-mode sweeping in a 650 nm--1030 nm astrocomb. \emph{Opt. Express} \textbf{2026}, \emph{34}, 8569--8578.

\bibitem{nc14-66}
Bai, B.; Yang, Q.; Shu, H.; et al. Microcomb-based integrated photonic processing unit. \emph{Nat. Commun.} \textbf{2023}, \emph{14}, 66.

\bibitem{nc16-292}
Bai, Y.; Xu, Y.; Chen, S.; et al. TOPS-speed complex-valued convolutional accelerator for feature extraction and inference. \emph{Nat. Commun.} \textbf{2025}, \emph{16}, 292.

\bibitem{eLight5-20}
Wang, Y.; Liao, K.; Zhang, K.; et al. Reconfigurable versatile integrated photonic computing chip. \emph{eLight} \textbf{2025}, \emph{5}, 20.

\bibitem{eLight5-10}
Yu, X.; Wei, Z.; Sha, F.; et al. Parallel optical computing capable of 100-wavelength multiplexing. \emph{eLight} \textbf{2025}, \emph{5}, 10.

\bibitem{sa11-eadt4252}
Wang, Y.; Wang, J.; Huang, J.; et al. Cross dual-microcomb dispersion interferometry ranging. \emph{Sci. Adv.} \textbf{2025}, \emph{11}, eadt4252.

\bibitem{sa11-eads9590}
Cai, Z.; Wang, Z.; Wei, Z.; et al. A microcomb-empowered Fourier domain mode-locked LIDAR. \emph{Sci. Adv.} \textbf{2025}, \emph{11}, eads9590.

\bibitem{nc13-3280}
Lukashchuk, A.; Riemensberger, J.; Karpov, M.; et al. Dual chirped microcomb based parallel ranging at megapixel-line rates. \emph{Nat. Commun.} \textbf{2022}, \emph{13}, 3280.

\bibitem{hansch}
Hänsch, T.W. Application of high-resolution laser spectroscopy. In \emph{Tunable Lasers and Applications}; Springer-Verlag: Berlin, Germany, 1976; pp. 326--339.

\bibitem{AP12-97}
Baklanov, Y.V.; Chebotayev, V.P. Narrow resonances of two-photon absorption of super-narrow pulses in a gas. \emph{Appl. Phys.} \textbf{1977}, \emph{12}, 97--99.

\bibitem{PRL38-760}
Teets, R.; Eckstein, J.; Hänsch, T.W. Coherent Two-Photon Excitation by Multiple Light Pulses. \emph{Phys. Rev. Lett.} \textbf{1977}, \emph{38}, 760--764.

\bibitem{PRL40-847}
Eckstein, J.N.; Ferguson, A.I.; Hänsch, T.W. High-Resolution Two-Photon Spectroscopy with Picosecond Light Pulses. \emph{Phys. Rev. Lett.} \textbf{1978}, \emph{40}, 847--850.

\bibitem{OL24-881}
Udem, T.; Reichert, J.; Holzwarth, R.; et al. Accurate measurement of large optical frequency differences with a mode-locked laser. \emph{Opt. Lett.} \textbf{1999}, \emph{24}, 881.

\bibitem{PRL82-3568}
Udem, T.; Reichert, J.; Holzwarth, R.; et al. Absolute Optical Frequency Measurement of the Cesium $D_{1}$ Line with a Mode-Locked Laser. \emph{Phys. Rev. Lett.} \textbf{1999}, \emph{82}, 3568--3571.

\bibitem{OC183-181}
Onae, A.; Ikegami, T.; Sugiyama, K.; et al. Optical frequency link between an acetylene stabilized laser at 1542 nm and an Rb stabilized laser at 778 nm using a two-color mode-locked fiber laser. \emph{Opt. Commun.} \textbf{2000}, \emph{183}, 181--187.

\bibitem{OE11-594}
Tauser, F.; Leitenstorfer, A.; Zinth, W. Amplified femtosecond pulses from an Er:fiber system: Nonlinear pulse shortening and self-referencing detection of the carrier-envelope phase evolution. \emph{Opt. Express} \textbf{2003}, \emph{11}, 594--600.

\bibitem{OE12-5872}
Adler, F.; Moutzouris, K.; Leitenstorfer, A.; et al. Phase-locked two-branch erbium-doped fiber laser system for long-term precision measurements of optical frequencies. \emph{Opt. Express} \textbf{2004}, \emph{12}, 5872--5880.

\bibitem{OE10-1404}
Rauschenberger, J.; Fortier, T.; Jones, D.; et al. Control of the frequency comb from a mode-locked Erbium-doped fiber laser. \emph{Opt. Express} \textbf{2002}, \emph{10}, 1404--1410.

\bibitem{science332-555}
Kippenberg, T.J.; Holzwarth, R.; Diddams, S.A. Microresonator-Based Optical Frequency Combs. \emph{Science} \textbf{2011}, \emph{332}, 555--559.

\bibitem{science361-eaan8083}
Kippenberg, T.J.; Gaeta, A.L.; Lipson, M.; et al. Dissipative Kerr solitons in optical microresonators. \emph{Science} \textbf{2018}, \emph{361}, eaan8083.

\bibitem{optica10-977}
Okawachi, Y.; Kim, B.Y.; Lipson, M.; et al. Chip-scale frequency combs for data communications in computing systems. \emph{Optica} \textbf{2023}, \emph{10}, 977--995.

\bibitem{aop18-86}
Sun, Y.; Wu, J.; Tan, M.; et al. Applications of optical microcombs. \emph{Adv. Opt. Photonics} \textbf{2023}, \emph{15}, 86--175.

\bibitem{npjnano1-26}
Zhang, X.; Wang, C.; Cheng, Z.; et al. Advances in resonator-based Kerr frequency combs with high conversion efficiencies. \emph{npj Nanophoton} \textbf{2024}, \emph{1}, 26.

\bibitem{aplphoton7-100901}
Hermans, A.; Van Gasse, K.; Kuyken, B. On-chip optical comb sources. \emph{APL Photonics} \textbf{2022}, \emph{7}, 100901.

\bibitem{PQE86-100437}
Nie, M.; Xie, Y.; Li, B.; et al. Photonic frequency microcombs based on dissipative Kerr and quadratic cavity solitons. \emph{Prog. Quantum Electron.} \textbf{2022}, \emph{86}, 100437.

\bibitem{OL47-1855}
Liu, K.; Jin, N.; Cheng, H.; et al. Ultralow 0.034 dB/m loss wafer-scale integrated photonics realizing 720 million Q and 380~{$\mu$}W threshold Brillouin lasing. \emph{Opt. Lett.} \textbf{2022}, \emph{47}, 1855.

\bibitem{nc16-8878}
Pan, T.Y.; Tan, T.; Duan, B.; et al. Boosting silica micro-rod $Q$ factor to $8.28 \times 10^{9}$ for fully stabilizing a soliton microcomb. \emph{Nat. Commun.} \textbf{2025}, \emph{16}, 8878.

\bibitem{optica11-1397}
Ji, X.; Wang, R.N.; Liu, Y.; et al. Efficient mass manufacturing of high-density, ultra-low-loss Si\textsubscript{3}N\textsubscript{4} photonic integrated circuits. \emph{Optica} \textbf{2024}, \emph{11}, 1397--1407.

\bibitem{COL20-032201}
Wan, S.; Niu, R.; Peng, J.L.; et al. Fabrication of the high-$Q$ $\mathrm{Si}_{3}\mathrm{N}_{4}$ microresonators for soliton microcombs. \emph{Chin. Opt. Lett.} \textbf{2022}, \emph{20}, 032201.

\bibitem{PLA137-393}
Braginsky, V.B.; Gorodetsky, M.L.; Ilchenko, V.S. Quality-factor and nonlinear properties of optical whispering-gallery modes. \emph{Phys. Lett. A} \textbf{1989}, \emph{137}, 393--397.

\bibitem{SCPMA65-104201}
Liu, J.; Bo, F.; Chang, L.; et al. Emerging material platforms for integrated microcavity photonics. \emph{Sci. China Phys. Mech. Astron.} \textbf{2022}, \emph{65}, 104201.

\bibitem{nc15-4192}
Ling, J.; Gao, Z.; Xue, S.; et al. Electrically empowered microcomb laser. \emph{Nat. Commun.} \textbf{2024}, \emph{15}, 4192.

\bibitem{nature605-457}
Shu, H.; Chang, L.; Tao, Y.; et al. Microcomb-driven silicon photonic systems. \emph{Nature} \textbf{2022}, \emph{605}, 457--463.

\bibitem{LPR20-e01659}
Wang, Y.; Wang, Z.; Xu, T.; et al. Compact turnkey soliton microcombs at microwave rates via wafer-scale fabrication. \emph{Laser Photonics Rev.} \textbf{2026}, \emph{20}, e01659.

\bibitem{nc15-7030}
Wildi, T.; Ulanov, A.E.; Voumard, T.; et al. Phase-stabilised self-injection-locked microcomb. \emph{Nat. Commun.} \textbf{2024}, \emph{15}, 7030.

\bibitem{nc16-4829}
Wan, S.; Wang, P.Y.; Li, M.; et al. Self-locked broadband Raman-electro-optic microcomb. \emph{Nat. Commun.} \textbf{2025}, \emph{16}, 4829.

\bibitem{PRL93-083904}
Kippenberg, T.J.; Spillane, S.M.; Vahala, K.J. Kerr-Nonlinearity Optical Parametric Oscillation in an Ultrahigh-Q Toroid Microcavity. \emph{Phys. Rev. Lett.} \textbf{2004}, \emph{93}, 083904.

\bibitem{PRL93-243905}
Savchenkov, A.A.; Matsko, A.B.; Strekalov, D.; et al. Low Threshold Optical Oscillations in a Whispering Gallery Mode CaF2 Resonator. \emph{Phys. Rev. Lett.} \textbf{2004}, \emph{93}, 243905.

\bibitem{nature450-1214}
Del’Haye, P.; Schliesser, A.; Arcizet, O.; et al. Optical frequency comb generation from a monolithic microresonator. \emph{Nature} \textbf{2007}, \emph{450}, 1214--1217.

\bibitem{PRL101-093902}
Savchenkov, A.A.; Matsko, A.B.; Ilchenko, V.S.; et al. Tunable Optical Frequency Comb with a Crystalline Whispering Gallery Mode Resonator. \emph{Phys. Rev. Lett.} \textbf{2008}, \emph{101}, 093902.

\bibitem{OL34-878}
Grudinin, I.S.; Yu, N.; Maleki, L. Generation of optical frequency combs with a CaF2 resonator. \emph{Opt. Lett.} \textbf{2009}, \emph{34}, 878--880.

\bibitem{PRL102-193902}
Braje, D.; Hollberg, L.; Diddams, S. Brillouin-Enhanced Hyperparametric Generation of an Optical Frequency Comb in a Monolithic Highly Nonlinear Fiber Cavity Pumped by a cw Laser. \emph{Phys. Rev. Lett.} \textbf{2009}, \emph{102}, 193902.

\bibitem{LSA12-33}
Xiao, Z.; Li, T.; Cai, M.; et al. Near-zero-dispersion soliton and broadband modulational instability Kerr microcombs in anomalous dispersion. \emph{Light Sci. Appl.} \textbf{2023}, \emph{12}, 33.

\bibitem{nc15-55}
Nie, M.; Musgrave, J.; Jia, K.; et al. Turnkey photonic flywheel in a microresonator-filtered laser. \emph{Nat. Commun.} \textbf{2024}, \emph{15}, 55.

\bibitem{OE17-16209}
Agha, I.H.; Okawachi, Y.; Gaeta, A.L. Theoretical and experimental investigation of broadband cascaded four-wave mixing in high-Q microspheres. \emph{Opt. Express} \textbf{2009}, \emph{17}, 16209--16215.

\bibitem{OL36-3398}
Okawachi, Y.; Saha, K.; Levy, J.S.; et al. Octave-spanning frequency comb generation in a silicon nitride chip. \emph{Opt. Lett.} \textbf{2011}, \emph{36}, 3398.

\bibitem{OL37-875}
Johnson, A.R.; Okawachi, Y.; Levy, J.S.; et al. Chip-based frequency combs with sub-100 GHz repetition rates. \emph{Opt. Lett.} \textbf{2012}, \emph{37}, 875--877.

\bibitem{nphoton5-770}
Ferdous, F.; Miao, H.; Leaird, D.E.; et al. Spectral line-by-line pulse shaping of on-chip microresonator frequency combs. \emph{Nat. Photonics} \textbf{2011}, \emph{5}, 770--776.

\bibitem{nphoton18-294}
Ulanov, A.E.; Wildi, T.; Pavlov, N.G.; et al. Synthetic reflection self-injection-locked microcombs. \emph{Nat. Photonics} \textbf{2024}, \emph{18}, 294--299.

\bibitem{nature646-843}
Ji, X.; Li, X.; Qiu, Z.; et al. Deterministic soliton microcombs in Cu-free photonic integrated circuits. \emph{Nature} \textbf{2025}, \emph{646}, 843--849.

\bibitem{OL36-2290}
Liang, W.; Savchenkov, A.A.; Matsko, A.B.; et al. Generation of near-infrared frequency combs from a $\mathrm{MgF}_2$ whispering gallery mode resonator. \emph{Opt. Lett.} \textbf{2011}, \emph{36}, 2290--2292.

\bibitem{nc4-1345}
Wang, C.Y.; Herr, T.; Del’Haye, P.; et al. Mid-infrared optical frequency combs at 2.5 {$\mu$}m based on crystalline microresonators. \emph{Nat. Commun.} \textbf{2013}, \emph{4}, 1345.

\bibitem{LPR18-2301329}
Fujii, S.; Wada, K.; Kogure, S.; et al. Mechanically Actuated Kerr Soliton Microcombs. \emph{Laser Photonics Rev.} \textbf{2024}, \emph{18}, 2301329.

\bibitem{nphoton6-480}
Herr, T.; Hartinger, K.; Riemensberger, J.; et al. Universal formation dynamics and noise of Kerr-frequency combs in microresonators. \emph{Nat. Photonics} \textbf{2012}, \emph{6}, 480--487.

\bibitem{nc14-4590}
Shen, B.; Shu, H.; Xie, W.; et al. Harnessing microcomb-based parallel chaos for random number generation and optical decision making. \emph{Nat. Commun.} \textbf{2023}, \emph{14}, 4590.

\bibitem{LSA13-66}
Li, P.; Li, Q.; Tang, W.; et al. Scalable parallel ultrafast optical random bit generation based on a single chaotic microcomb. \emph{Light Sci. Appl.} \textbf{2024}, \emph{13}, 66.

\bibitem{PRL58-2209}
Lugiato, L.A.; Lefever, R. Spatial Dissipative Structures in Passive Optical Systems. \emph{Phys. Rev. Lett.} \textbf{1987}, \emph{58}, 2209--2211.

\bibitem{PRA89-063814}
Godey, C.; Balakireva, I.V.; Coillet, A.; et al. Stability analysis of the spatiotemporal Lugiato-Lefever model for Kerr optical frequency combs in the anomalous and normal dispersion regimes. \emph{Phys. Rev. A} \textbf{2014}, \emph{89}, 063814.

\bibitem{OC91-401}
Haelterman, M.; Trillo, S.; Wabnitz, S. Dissipative modulation instability in a nonlinear dispersive ring cavity. \emph{Opt. Commun.} \textbf{1992}, \emph{91}, 401--407.

\bibitem{OL18-601}
Wabnitz, S. Suppression of interactions in a phase-locked soliton optical memory. \emph{Opt. Lett.} \textbf{1993}, \emph{18}, 601--603.

\bibitem{OL36-2845}
Matsko, A.B.; Savchenkov, A.A.; Liang, W.; et al. Mode-locked Kerr frequency combs. \emph{Opt. Lett.} \textbf{2011}, \emph{36}, 2845--2847.

\bibitem{PRL136-063801}
Wei, Z.; Suk, D.; Liu, C.; et al. Kerr-Induced Noise Quenching in Pulse Pumped Microcavity Solitons. \emph{Phys. Rev. Lett.} \textbf{2026}, \emph{136}, 063801.

\bibitem{nc14-1802}
Lao, C.; Jin, X.; Chang, L.; et al. Quantum decoherence of dark pulses in optical microresonators. \emph{Nat. Commun.} \textbf{2023}, \emph{14}, 1802.

\bibitem{nphoton4-471}
Leo, F.; Coen, S.; Kockaert, P.; et al. Temporal cavity solitons in one-dimensional Kerr media as bits in an all-optical buffer. \emph{Nat. Photonics} \textbf{2010}, \emph{4}, 471--476.

\bibitem{nphoton8-145}
Herr, T.; Brasch, V.; Jost, J.D.; et al. Temporal solitons in optical microresonators. \emph{Nat. Photonics} \textbf{2014}, \emph{8}, 145--152.

\bibitem{optica2-1078}
Yi, X.; Yang, Q.F.; Yang, K.Y.; et al. Soliton frequency comb at microwave rates in a high-Q silica microresonator. \emph{Optica} \textbf{2015}, \emph{2}, 1078--1085.

\bibitem{nc14-169}
Niu, R.; Li, M.; Wan, S.; et al. kHz-precision wavemeter based on reconfigurable microsoliton. \emph{Nat. Commun.} \textbf{2023}, \emph{14}, 169.

\bibitem{nc15-1661}
Zhang, M.; Ding, S.; Li, X.; et al. Strong interactions between solitons and background light in Brillouin-Kerr microcombs. \emph{Nat. Commun.} \textbf{2024}, \emph{15}, 1661.

\bibitem{science351-357}
Brasch, V.; Geiselmann, M.; Herr, T.; et al. Photonic chip--based optical frequency comb using soliton Cherenkov radiation. \emph{Science} \textbf{2016}, \emph{351}, 357--360.

\bibitem{nc10-680}
Raja, A.S.; Voloshin, A.S.; Guo, H.; et al. Electrically pumped photonic integrated soliton microcomb. \emph{Nat. Commun.} \textbf{2019}, \emph{10}, 680.

\bibitem{nphoton19-630}
Jin, X.; Xie, Z.; Zhang, X.; et al. Microresonator-referenced soliton microcombs with zeptosecond-level timing noise. \emph{Nat. Photonics} \textbf{2025}, \emph{19}, 630--636.

\bibitem{nature562-401}
Stern, B.; Ji, X.; Okawachi, Y.; et al. Battery-operated integrated frequency comb generator. \emph{Nature} \textbf{2018}, \emph{562}, 401--405.

\bibitem{LSA15-370}
Tang, J.; Yang, J.; Yan, E.; et al. Multimodal locking-enabled robust and day-scale low-repetition-rate soliton microcomb for high-precision metrology. \emph{Light Sci. Appl.} \textbf{2026}, \emph{15}, 370.

\bibitem{nphoton18-632}
Liu, Y.; Lao, C.; Wang, M.; et al. Integrated vortex soliton microcombs. \emph{Nat. Photonics} \textbf{2024}, \emph{18}, 632--637.

\bibitem{optica10-650}
Wildi, T.; Gaafar, M.A.; Voumard, T.; et al. Dissipative Kerr solitons in integrated Fabry--Perot microresonators. \emph{Optica} \textbf{2023}, \emph{10}, 650.

\bibitem{OL43-4366}
Gong, Z.; Bruch, A.; Shen, M.; et al. High-fidelity cavity soliton generation in crystalline AlN micro-ring resonators. \emph{Opt. Lett.} \textbf{2018}, \emph{43}, 4366.

\bibitem{OL47-746}
Gong, Z.; Bruch, A.W.; Yang, F.; et al. Quadratic strong coupling in AlN Kerr cavity solitons. \emph{Opt. Lett.} \textbf{2022}, \emph{47}, 746.

\bibitem{PR11-A10}
Liu, K.; Wang, Z.; Yao, S.; et al. Mitigating fast thermal instability by engineered laser sweep in AlN soliton microcomb generation. \emph{Photonics Res.} \textbf{2023}, \emph{11}, A10.

\bibitem{optica3-823}
Pu, M.; Ottaviano, L.; Semenova, E.; et al. Efficient frequency comb generation in AlGaAs-on-insulator. \emph{Optica} \textbf{2016}, \emph{3}, 823.

\bibitem{nphoton18-625}
Chen, B.; Zhou, Y.; Liu, Y.; et al. Integrated optical vortex microcomb. \emph{Nat. Photonics} \textbf{2024}, \emph{18}, 625--631.

\bibitem{OL48-3853}
Wu, L.; Xie, W.; Chen, H.J.; et al. AlGaAs soliton microcombs at room temperature. \emph{Opt. Lett.} \textbf{2023}, \emph{48}, 3853.

\bibitem{ncommun6-6299}
Griffith, A.G.; Lau, R.K.W.; Cardenas, J.; et al. Silicon-chip mid-infrared frequency comb generation. \emph{Nat. Commun.} \textbf{2015}, \emph{6}, 6299.

\bibitem{optica3-854}
Yu, M.; Okawachi, Y.; Griffith, A.G.; et al. Mode-locked mid-infrared frequency combs in a silicon microresonator. \emph{Optica} \textbf{2016}, \emph{3}, 854.

\bibitem{LPR17-2200219}
Xia, D.; Yang, Z.; Zeng, P.; et al. Integrated chalcogenide photonics for microresonator soliton combs. \emph{Laser Photonics Rev.} \textbf{2023}, \emph{17}, 2200219.

\bibitem{nc16-10133}
Xia, D.; Luo, L.; Wang, L.; et al. Reconfigurable chalcogenide integrated nonlinear photonics. \emph{Nat. Commun.} \textbf{2025}, \emph{16}, 10133.

\bibitem{optica11-1454}
Nardi, A.; Davydova, A.; Kuznetsov, N.; et al. Integrated chirped photonic-crystal cavities in gallium phosphide for broadband soliton generation. \emph{Optica} \textbf{2024}, \emph{11}, 1454.

\bibitem{nature629-784}
Wang, C.; Li, Z.; Riemensberger, J.; et al. Lithium tantalate photonic integrated circuits for volume manufacturing. \emph{Nature} \textbf{2024}, \emph{629}, 784--790.

\bibitem{PR13-1955}
Cai, J.; Wan, S.; Chen, B.; et al. Stable soliton microcomb generation in X-cut lithium tantalate via thermal-assisted photorefractive suppression. \emph{Photonics Res.} \textbf{2025}, \emph{13}, 1955--1963.

\bibitem{nc16-2389}
Lv, X.; Nie, B.; Yang, C.; et al. Broadband microwave-rate dark pulse microcombs in dissipation-engineered LiNbO\textsubscript{3} microresonators. \emph{Nat. Commun.} \textbf{2025}, \emph{16}, 2389.

\bibitem{LSA14-270}
Song, Y.; Hu, Y.; Lončar, M.; et al. Hybrid Kerr-electro-optic frequency combs on thin-film lithium niobate. \emph{Light Sci. Appl.} \textbf{2025}, \emph{14}, 270.

\bibitem{LSA13-225}
Song, Y.; Hu, Y.; Zhu, X.; et al. Octave-spanning Kerr soliton frequency combs in dispersion- and dissipation-engineered lithium niobate microresonators. \emph{Light Sci. Appl.} \textbf{2024}, \emph{13}, 225.

\bibitem{nc15-3921}
Cheng, R.; Yu, M.; Shams-Ansari, A.; et al. Frequency comb generation via synchronous pumped $\chi^{(3)}$ resonator on thin-film lithium niobate. \emph{Nat. Commun.} \textbf{2024}, \emph{15}, 3921.

\bibitem{eLight5-15}
Nie, B.; Lv, X.; Yang, C.; et al. Soliton microcombs in X-cut LiNbO\textsubscript{3} microresonators. \emph{eLight} \textbf{2025}, \emph{5}, 15.

\bibitem{LPR18-2300627}
Wan, S.; Wang, P.Y.; Ma, R.; et al. Photorefraction-assisted self-emergence of dissipative Kerr solitons. \emph{Laser Photonics Rev.} \textbf{2024}, \emph{18}, 2300627.

\bibitem{nphoton9-594}
Xue, X.; Xuan, Y.; Liu, Y.; et al. Mode-locked dark pulse Kerr combs in normal-dispersion microresonators. \emph{Nat. Photonics} \textbf{2015}, \emph{9}, 594--600.

\bibitem{OE23-7713}
Lobanov, V.E.; Lihachev, G.; Kippenberg, T.J.; et al. Frequency combs and platicons in optical microresonators with normal GVD. \emph{Opt. Express} \textbf{2015}, \emph{23}, 7713.

\bibitem{nc13-1771}
Lihachev, G.; Weng, W.; Liu, J.; et al. Platicon microcomb generation using laser self-injection locking. \emph{Nat. Commun.} \textbf{2022}, \emph{13}, 1771.

\bibitem{CP6-303}
Rebolledo-Salgado, I.; Quevedo-Galán, C.; Helgason, Ó.B.; et al. Platicon dynamics in photonic molecules. \emph{Commun. Phys.} \textbf{2023}, \emph{6}, 303.

\bibitem{nc13-3134}
Yu, S.P.; Lucas, E.; Zang, J.; et al. A continuum of bright and dark-pulse states in a photonic-crystal resonator. \emph{Nat. Commun.} \textbf{2022}, \emph{13}, 3134.

\bibitem{PRL128-033901}
Zhang, S.; Bi, T.; Ghalanos, G.N.; et al. Dark-Bright Soliton Bound States in a Microresonator. \emph{Phys. Rev. Lett.} \textbf{2022}, \emph{128}, 033901.

\bibitem{nphoton11-671}
Cole, D.C.; Lamb, E.S.; Del’Haye, P.; et al. Soliton crystals in Kerr resonators. \emph{Nat. Photonics} \textbf{2017}, \emph{11}, 671--676.

\bibitem{nc12-3179}
Lu, Z.; Chen, H.J.; Wang, W.; et al. Synthesized soliton crystals. \emph{Nat. Commun.} \textbf{2021}, \emph{12}, 3179.

\bibitem{nphys15-1071}
Karpov, M.; Pfeiffer, M.H.P.; Guo, H.; et al. Dynamics of soliton crystals in optical microresonators. \emph{Nat. Phys.} \textbf{2019}, \emph{15}, 1071--1077.

\bibitem{OE30-13690}
Li, J.; Wan, S.; Peng, J.L.; et al. Thermal tuning of mode crossing and the perfect soliton crystal in a Si\textsubscript{3}N\textsubscript{4} microresonator. \emph{Opt. Express} \textbf{2022}, \emph{30}, 13690.

\bibitem{LSA11-341}
Wang, C.; Li, J.; Yi, A.; et al. Soliton formation and spectral translation into visible on CMOS-compatible 4H-silicon-carbide-on-insulator platform. \emph{Light Sci. Appl.} \textbf{2022}, \emph{11}, 341.

\bibitem{LSA13-251}
Hu, F.; Vinod, A.K.; Wang, W.; et al. Spatio-temporal breather dynamics in microcomb soliton crystals. \emph{Light Sci. Appl.} \textbf{2024}, \emph{13}, 251.

\bibitem{nc13-848}
Taheri, H.; Matsko, A.B.; Maleki, L.; et al. All-optical dissipative discrete time crystals. \emph{Nat. Commun.} \textbf{2022}, \emph{13}, 848.

\bibitem{LPR19-2500257}
Taheri, H.; Matsko, A.B.; Wiesenfeld, K.A. Dynamics of Dissipative Cavity Solitons in a Lattice Trap. \emph{Laser Photonics Rev.} \textbf{2025}, \emph{19}, 2500257.

\bibitem{science383-1080}
Ji, Q.X.; Liu, P.; Jin, W.; et al. Multimodality integrated microresonators using the Moiré speedup effect. \emph{Science} \textbf{2024}, \emph{383}, 1080--1083.

\bibitem{optica10-279}
Ji, Q.X.; Jin, W.; Wu, L.; et al. Engineered zero-dispersion microcombs using CMOS-ready photonics. \emph{Optica} \textbf{2023}, \emph{10}, 279.

\bibitem{nc16-4780}
Liu, P.; Ji, Q.X.; Liu, J.Y.; et al. Near-visible integrated soliton microcombs with detectable repetition rates. \emph{Nat. Commun.} \textbf{2025}, \emph{16}, 4780.

\bibitem{nphoton15-305}
Helgason, Ó.B.; Arteaga-Sierra, F.R.; Ye, Z.; et al. Dissipative solitons in photonic molecules. \emph{Nat. Photonics} \textbf{2021}, \emph{15}, 305--310.

\bibitem{nphys17-604}
Tikan, A.; Riemensberger, J.; Komagata, K.; et al. Emergent nonlinear phenomena in a driven dissipative photonic dimer. \emph{Nat. Phys.} \textbf{2021}, \emph{17}, 604--610.

\bibitem{sa8-eabm6982}
Tikan, A.; Tusnin, A.; Riemensberger, J.; et al. Protected generation of dissipative Kerr solitons in supermodes of coupled optical microresonators. \emph{Sci. Adv.} \textbf{2022}, \emph{8}, eabm6982.

\bibitem{CP9-206}
Deshmukh, S.; Tusnin, A.; Tikan, A.; et al. Nonlinear periodic orbit solutions and their bifurcation structure at the origin of soliton hopping in coupled microresonators. \emph{Commun. Phys.} \textbf{2026}, \emph{9}, 206.

\bibitem{PRL134-123801}
Sanyal, S.; Okawachi, Y.; Zhao, Y.; et al. Nonlinear dynamics of coupled-resonator Kerr combs. \emph{Phys. Rev. Lett.} \textbf{2025}, \emph{134}, 123801.

\bibitem{optica11-940}
Gao, M.; Yuan, Z.; Yu, Y.; et al. Observation of interband Kelly sidebands in coupled-ring soliton microcombs. \emph{Optica} \textbf{2024}, \emph{11}, 940.

\bibitem{nphoton13-616}
Xue, X.; Zheng, X.; Zhou, B. Super-efficient temporal solitons in mutually coupled optical cavities. \emph{Nat. Photonics} \textbf{2019}, \emph{13}, 616--622.

\bibitem{nphoton17-992}
Helgason, Ó.B.; Girardi, M.; Ye, Z.; et al. Surpassing the nonlinear conversion efficiency of soliton microcombs. \emph{Nat. Photonics} \textbf{2023}, \emph{17}, 992--999.

\bibitem{LSA15-185}
Zhu, K.; Luo, X.; Wang, Y.; et al. Power-efficient ultra-broadband soliton microcombs in resonantly-coupled microresonators. \emph{Light Sci. Appl.} \textbf{2026}, \emph{15}, 185.

\bibitem{nphoton17-977}
Yuan, Z.; Gao, M.; Yu, Y.; et al. Soliton pulse pairs at multiple colours in normal dispersion microresonators. \emph{Nat. Photonics} \textbf{2023}, \emph{17}, 977--983.

\bibitem{LSA15-166}
Ji, Q.X.; Hou, H.; Ge, J.; et al. Multicolor interband solitons in microcombs. \emph{Light Sci. Appl.} \textbf{2026}, \emph{15}, 166.

\bibitem{PR12-2376}
Ghosh, A.; Pal, A.; Hill, L.; et al. Controlled light distribution with coupled microresonator chains via Kerr symmetry breaking. \emph{Photonics Res.} \textbf{2024}, \emph{12}, 2376.

\bibitem{OL51-1649}
Li, R.; Xu, L.; Imran, M.; et al. Symmetry-breaking bifurcation of coupled topological edge states. \emph{Opt. Lett.} \textbf{2026}, \emph{51}, 1649.

\bibitem{anderson}
Anderson, P.W. More Is Different: Broken symmetry and the nature of the hierarchical structure of science. \emph{Science} \textbf{1972}, \emph{177}, 393--396.

\bibitem{PRL49-405}
Thouless, D.J.; Kohmoto, M.; Nightingale, M.P.; et al. Quantized Hall Conductance in a Two-Dimensional Periodic Potential. \emph{Phys. Rev. Lett.} \textbf{1982}, \emph{49}, 405--408.

\bibitem{RMP82-3045}
Hasan, M.Z.; Kane, C.L. Colloquium: Topological insulators. \emph{Rev. Mod. Phys.} \textbf{2010}, \emph{82}, 3045--3067.

\bibitem{RMP83-1057}
Qi, X.L.; Zhang, S.C. Topological insulators and superconductors. \emph{Rev. Mod. Phys.} \textbf{2011}, \emph{83}, 1057--1110.

\bibitem{NRM7-196}
Wieder, B.J.; Bradlyn, B.; Cano, J.; et al. Topological materials discovery from crystal symmetry. \emph{Nat. Rev. Mater.} \textbf{2022}, \emph{7}, 196--216.

\bibitem{PRL95-226801}
Kane, C.L.; Mele, E.J. Quantum Spin Hall Effect in Graphene. \emph{Phys. Rev. Lett.} \textbf{2005}, \emph{95}, 226801.

\bibitem{science318-766}
König, M.; Wiedmann, S.; Brüne, C.; et al. Quantum Spin Hall Insulator State in HgTe Quantum Wells. \emph{Science} \textbf{2007}, \emph{318}, 766--770.

\bibitem{RMP89-040502}
Haldane, F.D.M. Nobel Lecture: Topological quantum matter. \emph{Rev. Mod. Phys.} \textbf{2017}, \emph{89}, 040502.

\bibitem{RMP89-040501}
Kosterlitz, J.M. Nobel Lecture: Topological defects and phase transitions. \emph{Rev. Mod. Phys.} \textbf{2017}, \emph{89}, 040501.

\bibitem{PRL100-013904}
Haldane, F.D.M.; Raghu, S. Possible Realization of Directional Optical Waveguides in Photonic Crystals with Broken Time-Reversal Symmetry. \emph{Phys. Rev. Lett.} \textbf{2008}, \emph{100}, 013904.

\bibitem{PRL100-013905}
Wang, Z.; Chong, Y.D.; Joannopoulos, J.D.; et al. Reflection-Free One-Way Edge Modes in a Gyromagnetic Photonic Crystal. \emph{Phys. Rev. Lett.} \textbf{2008}, \emph{100}, 013905.

\bibitem{nature461-772}
Wang, Z.; Chong, Y.; Joannopoulos, J.D.; et al. Observation of unidirectional backscattering-immune topological electromagnetic states. \emph{Nature} \textbf{2009}, \emph{461}, 772--775.

\bibitem{RMP91-015006}
Ozawa, T.; Price, H.M.; Amo, A.; et al. Topological Photonics. \emph{Rev. Mod. Phys.} \textbf{2019}, \emph{91}, 015006.

\bibitem{LSA9-1}
Kim, M.; Jacob, Z.; Rho, J. Recent advances in 2D, 3D and higher-order topological photonics. \emph{Light Sci. Appl.} \textbf{2020}, \emph{9}, 130.

\bibitem{APR}
Xue, H.; Yang, Y.; Zhang, B. Topological Valley Photonics: Physics and Device Applications. \emph{Adv. Photonics Res.} \textbf{2021}, \emph{2},~2100013.

\bibitem{nc15-931}
Khanikaev, A.B.; Alù, A. Topological photonics: robustness and beyond. \emph{Nat. Commun.} \textbf{2024}, \emph{15}, 931.

\bibitem{NRP8-327}
Ma, A.; Dai, T.; Li, G.; et al. Reconfigurable and programmable integrated topological photonics. \emph{Nat. Rev. Phys.} \textbf{2026}, \emph{8}, 327--343.

\bibitem{nphys7-907}
Hafezi, M.; Demler, E.A.; Lukin, M.D.; et al. Robust optical delay lines with topological protection. \emph{Nat. Phys.} \textbf{2011}, \emph{7}, 907--912.

\bibitem{nphoton7-1001}
Hafezi, M.; Mittal, S.; Fan, J.; et al. Imaging topological edge states in silicon photonics. \emph{Nat. Photonics} \textbf{2013}, \emph{7}, 1001--1005.

\bibitem{np6-782}
Fang, K.; Yu, Z.; Fan, S. Realizing effective magnetic field for photons by controlling the phase of dynamic modulation. \emph{Nat. Photonics} \textbf{2012}, \emph{6}, 782--787.

\bibitem{nmater23-928}
Dai, T.; Ma, A.; Mao, J.; et al. A programmable topological photonic chip. \emph{Nat. Mater.} \textbf{2024}, \emph{23}, 928--936.

\bibitem{nmater12-233}
Khanikaev, A.B.; Mousavi, S.H.; Tse, W.K.; et al. Photonic topological insulators. \emph{Nat. Mater.} \textbf{2013}, \emph{12}, 233--239.

\bibitem{nmater15-542}
Cheng, X.; Jouvaud, C.; Ni, X.; et al. Robust reconfigurable electromagnetic pathways within a photonic topological insulator. \emph{Nat. Mater.} \textbf{2016}, \emph{15}, 542--548.

\bibitem{nc5-5782}
Chen, W.J.; Jiang, S.J.; Chen, X.D.; et al. Experimental realization of photonic topological insulator in a uniaxial metacrystal waveguide. \emph{Nat. Commun.} \textbf{2014}, \emph{5}, 5782.

\bibitem{CR123-7585}
Ni, X.; Yves, S.; Krasnok, A.; et al. Topological Metamaterials. \emph{Chem. Rev.} \textbf{2023}, \emph{123}, 7585--7654.

\bibitem{nature496-196}
Rechtsman, M.C.; Zeuner, J.M.; Plotnik, Y.; et al. Photonic Floquet topological insulators. \emph{Nature} \textbf{2013}, \emph{496}, 196--200.

\bibitem{nc17-3020}
Jiang, T.; Tian, Z.N.; Tao, R.; et al. Photonic non-Abelian topological insulators with six bands. \emph{Nat. Commun.} \textbf{2026}, \emph{17}, 3020.

\bibitem{nc17-966}
Wu, J.L.; Xiao, K.H.; Ni, X.; et al. Superadiabatic topological pumping on photonic chips. \emph{Nat. Commun.} \textbf{2026}, \emph{17}, 966.

\bibitem{nc15-9311}
Sun, Y.K.; Shan, Z.L.; Tian, Z.N.; et al. Two-dimensional non-Abelian Thouless pump. \emph{Nat. Commun.} \textbf{2024}, \emph{15}, 9311.

\bibitem{nmater16-298}
Dong, J.W.; Chen, X.D.; Zhu, H.; et al. Valley photonic crystals for control of spin and topology. \emph{Nat. Mater.} \textbf{2017}, \emph{16}, 298--302.

\bibitem{nphys14-140}
Gao, F.; Xue, H.; Yang, Z.; et al. Topologically protected refraction of robust kink states in valley photonic crystals. \emph{Nat. Phys.} \textbf{2018}, \emph{14}, 140--144.

\bibitem{LPR16-2100300}
Tang, G.J.; He, X.T.; Shi, F.L.; et al. Topological Photonic Crystals: Physics, Designs, and Applications. \emph{Laser Photonics Rev.} \textbf{2022}, \emph{16}, 2100300.

\bibitem{PNAS121-e2411793121}
Yang, K.; Fu, Q.; Prates, H.C.; et al. Observation of Thouless pumping of light in quasiperiodic photonic crystals. \emph{Proc. Natl. Acad. Sci. USA} \textbf{2024}, \emph{121}, e2411793121.

\bibitem{nc13-6738}
Wang, P.; Fu, Q.; Peng, R.; et al. Two-dimensional Thouless pumping of light in photonic moiré lattices. \emph{Nat. Commun.} \textbf{2022}, \emph{13}, 6738.

\bibitem{APR7-021306}
Smirnova, D.; Leykam, D.; Chong, Y.; et al. Nonlinear topological photonics. \emph{Appl. Phys. Rev.} \textbf{2020}, \emph{7}, 021306.

\bibitem{nphys20-905}
Szameit, A.; Rechtsman, M.C. Discrete nonlinear topological photonics. \emph{Nat. Phys.} \textbf{2024}, \emph{20}, 905--912.

\bibitem{PRA113-053506}
Li, R.; Imran, M.; Wang, W.; et al. Dark solitons in nonlinear Su-Schrieffer-Heeger lattices. \emph{Phys. Rev. A} \textbf{2026}, \emph{113}, 053506.

\bibitem{CP5-275}
Li, R.; Kong, X.; Hang, D.; et al. Topological bulk solitons in a nonlinear photonic Chern insulator. \emph{Commun. Phys.} \textbf{2022}, \emph{5}, 275.

\bibitem{PRB105-L201111}
Li, R.; Li, P.; Jia, Y.; et al. Self-localized topological states in three dimensions. \emph{Phys. Rev. B} \textbf{2022}, \emph{105}, L201111.

\bibitem{AP12-2291}
Wong, S.; Cerjan, A.; Makris, K.G.; et al. Nonlinear topological photonics: Capturing nonlinear dynamics and optical thermodynamics. \emph{ACS Photonics} \textbf{2025}, \emph{12}, 2291--2303.

\bibitem{RMP83-247}
Kartashov, Y.V.; Malomed, B.A.; Torner, L. Solitons in nonlinear lattices. \emph{Rev. Mod. Phys.} \textbf{2011}, \emph{83}, 247--305.

\bibitem{PRA90-023813}
Ablowitz, M.J.; Curtis, C.W.; Ma, Y.P. Linear and nonlinear traveling edge waves in optical honeycomb lattices. \emph{Phys. Rev. A} \textbf{2014}, \emph{90}, 023813.

\bibitem{PRA94-021801}
Lumer, Y.; Rechtsman, M.C.; Plotnik, Y.; et al. Instability of bosonic topological edge states in the presence of interactions. \emph{Phys. Rev. A} \textbf{2016}, \emph{94}, 021801.

\bibitem{PRL128-093901}
Kartashov, Y.V.; Arkhipova, A.A.; Zhuravitskii, S.A.; et al. Observation of Edge Solitons in Topological Trimer Arrays. \emph{Phys. Rev. Lett.} \textbf{2022}, \emph{128}, 093901.

\bibitem{PRL117-143901}
Leykam, D.; Chong, Y.D. Edge Solitons in Nonlinear-Photonic Topological Insulators. \emph{Phys. Rev. Lett.} \textbf{2016}, \emph{117}, 143901.

\bibitem{PRX11-041057}
Mukherjee, S.; Rechtsman, M.C. Observation of Unidirectional Solitonlike Edge States in Nonlinear Floquet Topological Insulators. \emph{Phys. Rev. X} \textbf{2021}, \emph{11}, 041057.

\bibitem{PRL111-243905}
Lumer, Y.; Plotnik, Y.; Rechtsman, M.C.; et al. Self-Localized States in Photonic Topological Insulators. \emph{Phys. Rev. Lett.} \textbf{2013}, \emph{111}, 243905.

\bibitem{science368-856}
Mukherjee, S.; Rechtsman, M.C. Observation of Floquet solitons in a topological bandgap. \emph{Science} \textbf{2020}, \emph{368}, 856--859.

\bibitem{nphys17-995}
Kirsch, M.S.; Zhang, Y.; Kremer, M.; et al. Nonlinear second-order photonic topological insulators. \emph{Nat. Phys.} \textbf{2021}, \emph{17}, 995--1000.

\bibitem{LSA13-264}
Zhong, H.; Kompanets, V.O.; Zhang, Y.; et al. Observation of nonlinear fractal higher order topological insulator. \emph{Light Sci. Appl.} \textbf{2024}, \emph{13}, 264.

\bibitem{LSA12-194}
Ren, B.; Arkhipova, A.A.; Zhang, Y.; et al. Observation of nonlinear disclination states. \emph{Light Sci. Appl.} \textbf{2023}, \emph{12}, 194.

\bibitem{LSA9-147}
Xia, S.; Jukić, D.; Wang, N.; et al. Nontrivial coupling of light into a defect: the interplay of nonlinearity and topology. \emph{Light Sci. Appl.} \textbf{2020}, \emph{9}, 147.

\bibitem{science372-72}
Xia, S.; Kaltsas, D.; Song, D.; et al. Nonlinear tuning of PT symmetry and non-Hermitian topological states. \emph{Science} \textbf{2021}, \emph{372}, 72--76.

\bibitem{LSA10-164}
Hu, Z.; Bongiovanni, D.; Jukić, D.; et al. Nonlinear control of photonic higher-order topological bound states in the continuum. \emph{Light Sci. Appl.} \textbf{2021}, \emph{10}, 164.

\bibitem{nanophotonics14-769}
Jörg, C.; Jürgensen, M.; Mukherjee, S.; et al. Optical control of topological end states via soliton formation in a 1D lattice. \emph{Nanophotonics} \textbf{2025}, \emph{14}, 769--775.

\bibitem{PRL127-184101}
Bongiovanni, D.; Jukić, D.; Hu, Z.; et al. Dynamically Emerging Topological Phase Transitions in Nonlinear Interacting Soliton Lattices. \emph{Phys. Rev. Lett.} \textbf{2021}, \emph{127}, 184101.

\bibitem{nature596-63}
J\"{u}rgensen, M.; Mukherjee, S.; Rechtsman, M.C. Quantized nonlinear Thouless pumping. \emph{Nature} \textbf{2021}, \emph{596}, 63--67.

\bibitem{PRL128-154101}
Fu, Q.; Wang, P.; Kartashov, Y.V.; et al. Nonlinear Thouless Pumping: Solitons and Transport Breakdown. \emph{Phys. Rev. Lett.} \textbf{2022}, \emph{128}, 154101.

\bibitem{PRL128-113901}
Jürgensen, M.; Rechtsman, M.C. Chern Number Governs Soliton Motion in Nonlinear Thouless Pumps. \emph{Phys. Rev. Lett.} \textbf{2022}, \emph{128}, 113901.

\bibitem{ncommun13-5997}
Mostaan, N.; Grusdt, F.; Goldman, N. Quantized topological pumping of solitons in nonlinear photonics and ultracold atomic mixtures. \emph{Nat. Commun.} \textbf{2022}, \emph{13}, 5997.

\bibitem{nphys19-420}
Jürgensen, M.; Mukherjee, S.; Jörg, C.; et al. Quantized fractional Thouless pumping of solitons. \emph{Nat. Phys.} \textbf{2023}, \emph{19}, 420--426.

\bibitem{CP8-360}
Cai, Z.F.; Wang, Y.C.; Zhang, Y.R.; et al. Versatile control of nonlinear topological states in non-Hermitian systems. \emph{Commun. Phys.} \textbf{2025}, \emph{8}, 360.

\bibitem{nature561-502}
Mittal, S.; Goldschmidt, E.A.; Hafezi, M. A topological source of quantum light. \emph{Nature} \textbf{2018}, \emph{561}, 502--506.

\bibitem{nphoton16-248}
Dai, T.; Ao, Y.; Bao, J.; et al. Topologically protected quantum entanglement emitters. \emph{Nat. Photonics} \textbf{2022}, \emph{16}, 248--257.

\bibitem{nphoton15-542}
Mittal, S.; Orre, V.V.; Goldschmidt, E.A.; et al. Tunable quantum interference using a topological source of indistinguishable photon pairs. \emph{Nat. Photonics} \textbf{2021}, \emph{15}, 542--548.

\bibitem{nnano14-2}
Kruk, S.; Poddubny, A.; Smirnova, D.; et al. Nonlinear light generation in topological nanostructures. \emph{Nat. Nanotechnol.} \textbf{2019}, \emph{14}, 126--130.

\bibitem{OL42-5174}
Pilozzi, L.; Conti, C. Topological cascade laser for frequency comb generation in $\mathcal{PT}$-symmetric structures. \emph{Opt. Lett.} \textbf{2017}, \emph{42}, 5174.

\bibitem{nphys17-1169}
Mittal, S.; Moille, G.; Srinivasan, K.; et al. Topological frequency combs and nested temporal solitons. \emph{Nat. Phys.} \textbf{2021}, \emph{17}, 1169--1176.

\bibitem{nphys17-1078}
Peano, V. Matryoshka frequency combs. \emph{Nat. Phys.} \textbf{2021}, \emph{17}, 1078--1079.

\bibitem{arxiv-OFC}
Englebert, N.; Gray, R.M.; Ledezma, L.; et al. Topological soliton frequency comb in nanophotonic lithium niobate. \emph{Nature} \textbf{2026}, \emph{652}, 76--81.

\bibitem{nphoton20-616}
Englebert, N.; Gray, R.M.; Marandi, A.; et al. Temporal solitons and frequency combs in quadratic resonators. \emph{Nat. Photonics} \textbf{2026}, \emph{20}, 616--627.

\bibitem{science384-1356}
Flower, C.J.; Jalali Mehrabad, M.; Xu, L.; et al. Observation of topological frequency combs. \emph{Science} \textbf{2024}, \emph{384}, 1356--1361.

\bibitem{cp6-317}
Tusnin, A.; Tikan, A.; Komagata, K.; et al. Nonlinear dynamics and Kerr frequency comb formation in lattices of coupled microresonators. \emph{Commun. Phys.} \textbf{2023}, \emph{6}, 317.

\bibitem{PRB108-205421}
Jiang, Z.; Zhou, L.; Li, W.; et al. Topological dissipative Kerr soliton combs in a valley photonic crystal resonator. \emph{Phys. Rev. B} \textbf{2023}, \emph{108}, 205421.

\bibitem{AQT}
Jiang, Z.; Chen, Y.; Jiang, C.; et al. Generation of Quantum Optical Frequency Combs in Topological Resonators. \emph{Adv. Quantum Technol.} \textbf{2024}, \emph{7}, 2300354.

\bibitem{PR13-163}
Jiang, Z.; Wang, H.; Xie, P.; et al. On-chip topological transport of integrated optical frequency combs. \emph{Photonics Res.} \textbf{2025}, \emph{13}, 163.

\bibitem{AP12-219}
Huang, L.; Zhang, W.; Zong, Y.; et al. Hyperbolic Topological Frequency Combs. \emph{ACS Photonics} \textbf{2025}, \emph{12}, 219--226.

\bibitem{nc15-9642}
Hashemi, S.D.; Mittal, S. Floquet topological dissipative Kerr solitons and incommensurate frequency combs. \emph{Nat. Commun.} \textbf{2024}, \emph{15}, 9642.

\bibitem{sa11-eadu6554}
Hashemi, S.D.; Mittal, S. Reconfigurable non-Hermitian soliton combs using dissipative couplings and topological windings. \emph{Sci. Adv.} \textbf{2025}, \emph{11}, eadu6554.

\bibitem{arxiv2}
Hashemi, S.D.; Maisuriya, A.; Mittal, S. Collective Topological Dark Solitons, Platicons, and Bright-Like Solitons in Normal Dispersion Optical Frequency Combs. \emph{arXiv} \textbf{2026}, arXiv:2608.23708.

\bibitem{nphoton17-937}
Blanco-Redondo, A.; de Sterke, C.M.; Xu, C.; et al. The bright prospects of optical solitons after 50 years. \emph{Nat. Photonics} \textbf{2023}, \emph{17}, 937--942.

\bibitem{optica10-1452}
Choi, J.; Sohn, B.U.; Sahin, E.; et al. Pure-quartic Bragg solitons in chip-scale nonlinear integrated circuits. \emph{Optica} \textbf{2023}, \emph{10}, 1452.

\bibitem{nphoton17-943}
Lucas, E.; Yu, S.P.; Briles, T.C.; et al. Tailoring microcombs with inverse-designed, meta-dispersion microresonators. \emph{Nat. Photonics} \textbf{2023}, \emph{17}, 943--950.

\bibitem{nc13-4764}
Anderson, M.H.; Weng, W.; Lihachev, G.; et al. Zero dispersion Kerr solitons in optical microresonators. \emph{Nat. Commun.} \textbf{2022}, \emph{13}, 4764.

\bibitem{CP6-144}
Moille, G.; Lu, X.; Stone, J.; et al. Fourier synthesis dispersion engineering of photonic crystal microrings for broadband frequency combs. \emph{Commun. Phys.} \textbf{2023}, \emph{6}, 144.

\bibitem{OL50-2073}
Silvestri, C.; Widjaja, J.; Lin, A.; et al. Theory of multicolor soliton microcombs. \emph{Opt. Lett.} \textbf{2025}, \emph{50}, 2073.

\bibitem{OL50-4262}
Silvestri, C.; Long Qiang, Y.; Panda, K.; et al. Pure high-order dispersion dissipative Kerr solitons in optical cavities. \emph{Opt. Lett.} \textbf{2025}, \emph{50}, 4262.

\bibitem{OL47-2438}
Parra-Rivas, P.; Hetzel, S.; Kartashov, Y.V.; et al. Quartic Kerr cavity combs: bright and dark solitons. \emph{Opt. Lett.} \textbf{2022}, \emph{47}, 2438.

\bibitem{nature618-687}
Zhang, X.; Zangeneh-Nejad, F.; Chen, Z.G.; et al. A second wave of topological phenomena in photonics and acoustics. \emph{Nature} \textbf{2023}, \emph{618}, 687--697.

\bibitem{science383-eadf9621}
Yang, Y.; Yang, B.; Ma, G.; et al. Non-Abelian physics in light and sound. \emph{Science} \textbf{2024}, \emph{383}, eadf9621.

\bibitem{FOP20-44204}
Li, R.; Wang, W.; Kong, X.; et al. Realization of a non-Hermitian Haldane model in circuits. \emph{Front. Phys.} \textbf{2025}, \emph{20}, 044204.

\bibitem{NMI6-904}
Long, Y.; Xue, H.; Zhang, B. Unsupervised learning of topological non-Abelian braiding in non-Hermitian bands. \emph{Nat. Mach. Intell.} \textbf{2024}, \emph{6}, 904--910.

\bibitem{PRB113-085413}
Qin, X.; Zhang, W.; Cao, W.; et al. Global Floquet band braiding in non-Hermitian space-time crystals. \emph{Phys. Rev. B} \textbf{2026}, \emph{113}, 085413.

\bibitem{PRL130-157201}
Guo, C.X.; Chen, S.; Ding, K.; et al. Exceptional non-Abelian topology in multiband non-Hermitian systems. \emph{Phys. Rev. Lett.} \textbf{2023}, \emph{130}, 157201.

\bibitem{nc15-1647}
Huang, L.; He, L.; Zhang, W.; et al. Hyperbolic photonic topological insulators. \emph{Nat. Commun.} \textbf{2024}, \emph{15}, 1647.

\bibitem{PRL129-246402}
Urwyler, D.M.; Lenggenhager, P.M.; Boettcher, I.; et al. Hyperbolic Topological Band Insulators. \emph{Phys. Rev. Lett.} \textbf{2022}, \emph{129}, 246402.

\bibitem{nc13-2937}
Zhang, W.; Yuan, H.; Sun, N.; et al. Observation of novel topological states in hyperbolic lattices. \emph{Nat. Commun.} \textbf{2022}, \emph{13}, 2937.

\bibitem{nc15-2293}
Chen, Q.; Zhang, Z.; Qin, H.; et al. Anomalous and Chern topological waves in hyperbolic networks. \emph{Nat. Commun.} \textbf{2024}, \emph{15}, 2293.

\bibitem{science357-61}
Benalcazar, W.A.; Bernevig, B.A.; Hughes, T.L. Quantized electric multipole insulators. \emph{Science} \textbf{2017}, \emph{357}, 61--66.

\bibitem{PRL120-026801}
Ezawa, M. Higher-Order Topological Insulators and Semimetals on the Breathing Kagome and Pyrochlore Lattices. \emph{Phys. Rev. Lett.} \textbf{2018}, \emph{120}, 026801.

\bibitem{CP8-451}
Huang, C.; Kireev, A.V.; Jiang, Y.; et al. Observation of nonlinear higher-order topological insulators with unconventional boundary truncations. \emph{Commun. Phys.} \textbf{2025}, \emph{8}, 451.

\bibitem{CSF207-118044}
Li, R.; Wang, W.; Jia, Y.; et al. Nonlinear quadrupole topological insulators. \emph{Chaos Solitons Fractals} \textbf{2026}, \emph{207}, 118044.

\bibitem{nphys17-704}
Liu, Y.G.N.; Jung, P.S.; Parto, M.; et al. Gain-induced topological response via tailored long-range interactions. \emph{Nat. Phys.} \textbf{2021}, \emph{17}, 704--709.

\bibitem{science370-701}
Maczewsky, L.J.; Heinrich, M.; Kremer, M.; et al. Nonlinearity-induced photonic topological insulator. \emph{Science} \textbf{2020}, \emph{370}, 701--704.

\bibitem{nc13-3379}
Zhou, D.; Rocklin, D.Z.; Leamy, M.; et al. Topological invariant and anomalous edge modes of strongly nonlinear systems. \emph{Nat. Commun.} \textbf{2022}, \emph{13}, 3379.

\bibitem{nc16-422}
Sone, K.; Ezawa, M.; Gong, Z.; et al. Transition from the topological to the chaotic in the nonlinear Su--Schrieffer--Heeger model. \emph{Nat. Commun.} \textbf{2025}, \emph{16}, 422.

\bibitem{nphys20-1164}
Sone, K.; Ezawa, M.; Ashida, Y.; et al. Nonlinearity-induced topological phase transition characterized by the nonlinear Chern number. \emph{Nat. Phys.} \textbf{2024}, \emph{20}, 1164--1170.

\bibitem{PRB113-214310}
Li, R.; Wang, W.; Kong, X.; et al. Nonlinear topological edge states, topological gap solitons, and self-induced topological edge states in nonlinear Su-Schrieffer-Heeger circuit lattices. \emph{Phys. Rev. B} \textbf{2026}, \emph{113}, 214310.

\bibitem{nc17-6718}
Tao, Y.L.; Wang, J.H.; Xu, Y. Anomalous quantized nonlinear soliton pumping. \emph{Nat. Commun.} \textbf{2026}, \emph{17}, 6718.

\bibitem{NRP8-55}
Leykam, D.; Xue, H.; Zhang, B.; et al. Limitations and possibilities of topological photonics. \emph{Nat. Rev. Phys.} \textbf{2026}, \emph{8}, 55--64.

\bibitem{NC12-1092}
Karnieli, A.; Tsesses, S.; Bartal, G.; et al. Emulating spin transport with nonlinear optics, from high-order skyrmions to the topological Hall effect. \emph{Nat. Commun.} \textbf{2021}, \emph{12}, 1092.
\end{thebibliography}
\end{document}